\documentclass{aa}  
\makeatletter
\renewcommand*\aa@pageof{, page \thepage{} of \pageref*{LastPage}}
\usepackage{natbib}
\bibpunct{(}{)}{;}{a}{}{,} % to follow the A&A style

\usepackage{enumitem}

\usepackage{graphicx}
\usepackage{txfonts}
\usepackage{booktabs}

\usepackage[colorlinks=true,allcolors=blue]{hyperref}
\usepackage{booktabs}

\begin{document}

   \title{Beyond the horizon: Quantifying the full sky foreground wedge in the cylindrical power spectrum}

   \subtitle{}

   \author{S. Munshi\inst{1}
          \and
          F. G. Mertens\inst{2,1}
          \and
          L. V. E. Koopmans\inst{1}
          \and
          A. R. Offringa\inst{3,1}
          \and
          E. Ceccotti\inst{1}
          \and
          S. A. Brackenhoff\inst{1}
          \and
          J. K. Chege\inst{1}
          \and
          B. K. Gehlot\inst{1}
          \and
          S. Ghosh\inst{1}
          \and
          C. H\"{o}fer\inst{1}
          \and
          M. Mevius\inst{3}
          }

   \institute{Kapteyn Astronomical Institute, University of Groningen, P.O. Box 800, 9700 AV Groningen, The Netherlands\\ \email{munshi@astro.rug.nl}
         \and
    LERMA, Observatoire de Paris, Universit\'{e} PSL, CNRS, Sorbonne Universit\'{e}, F-75014 Paris, France
        \and
    ASTRON, PO Box 2, 7990 AA Dwingeloo, The Netherlands
    }

   \date{Received --; accepted --}
 
  \abstract
    {One of the main obstacles preventing the detection of the redshifted 21-cm signal from neutral hydrogen in the early Universe is the astrophysical foreground emission, which is several orders of magnitude brighter than the signal. The foregrounds, due to their smooth spectra, are expected to predominantly occupy a region in the cylindrical power spectrum known as the foreground wedge. However, the conventional equations describing the extent of the foreground wedge are derived under a flat-sky approximation. This assumption breaks down for tracking wide-field instruments, thus rendering these equations inapplicable in these situations. In this paper, we derive equations for the full sky foreground wedge and show that the foregrounds can potentially extend far beyond what the conventional equations suggest. We also derive the equations that describe a specific bright source in the cylindrical power spectrum space. The validity of both sets of equations is tested against numerical simulations. Many current and upcoming interferometers (e.g., LOFAR, NenuFAR, MWA, SKA) are wide-field phase-tracking instruments. These equations give us new insights into the nature of foreground contamination in the cylindrical power spectra estimated using wide-field instruments. Additionally, they allow us to accurately associate features in the power spectrum to foregrounds or instrumental effects. The equations are also important for correctly selecting the ``EoR window'' for foreground avoidance analyses, and for planning 21-cm observations. In future analyses, it is recommended to use these updated horizon lines to indicate the foreground wedge in the cylindrical power spectrum accurately. The new equations for generating the updated wedge lines are made available in a Python library, \texttt{pslines}.}


   \maketitle
%
%________________________________________________________________

\section{Introduction}\label{sec:introduction}
One of the most comprehensive probes of the astrophysics and cosmology of the early Universe is the redshifted 21-cm signal from neutral hydrogen (HI) \citep{madau199721,shaver1999can}. The fluctuations in the brightness temperature distribution of the 21-cm signal against the cosmic microwave background radiation at high redshifts (e.g., $z>6$) can be probed by low-frequency radio interferometers. However, none of the currently operating instruments have the thermal noise sensitivity to detect the faint signal in tomographic images in reasonable observing time (i.e., thousands of hours). Hence they attempt to detect the 21-cm signal statistically and quantify it in terms of a power spectrum. Several telescopes such as the GMRT\footnote{Giant Metrewave Radio Telescope, \url{http://www.gmrt.ncra.tifr.res.in}} \citep{paciga2013simulation}, MWA\footnote{Murchison Widefield Array, \url{https://www.mwatelescope.org}} \citep{ewall2016first,barry2019improving,li2019first,trott2020deep,yoshiura2021new}, LOFAR\footnote{Low-Frequency Array, \url{http://www.lofar.org}} \citep{patil2017upper,gehlot2019first,gehlot2020aartfaac,mertens2020improved}, PAPER\footnote{Precision Array to Probe EoR} \citep{kolopanis2019simplified}, HERA\footnote{Hydrogen Epoch of Reionization Array, \url{https://reionization.org}} \citep{abdurashidova2022first,adams2023improved}, OVRO-LWA\footnote{Owen's Valley Radio Observatory - Long Wavelength Array} \citep{eastwood201921,garsden202121}, and NenuFAR\footnote{New Extension in Nan\c cay Upgrading LOFAR, \url{https://nenufar.obs-nancay.fr}} \citep{munshi2024first} have already set upper limits on the 21-cm signal power spectrum during the Cosmic Dawn (CD) and the subsequent Epoch of Reionization (EoR). The future SKA\footnote{Square Kilometer Array \url{https://www.skao.int}}\citep{dewdney2009square, koopmans2015cosmic} is expected to have the sensitivity to generate tomographic maps of the 21-cm signal in the early Universe. A detection or strong upper limits will allow us to constrain the underlying astrophysical models that cause the fluctuations of the 21-cm signal during these epochs \citep[e.g.,][]{furlanetto2006cosmology,pritchard201221}. However, the orders of magnitude brighter foregrounds and instrumental systematics are the two most important obstacles to the detection of the power spectrum of the 21-cm signal.

A crucial diagnostic in any 21-cm cosmology analysis is the cylindrical power spectrum in which the signal brightness temperature cube is averaged within cylindrical shells in the Fourier space, with the axis of the cylinders pointing along the line of sight (LoS). The cylindrical power spectrum is an estimator of the variance in the observed data as a function of the spatial mode along the LoS ($k_{\parallel}$) and the spatial mode on the plane perpendicular to the LoS ($k_{\perp}$). The intrinsic chromatic nature of an interferometer causes the spatial scale probed by a certain baseline to vary with frequency. This leads to a mixture of spatial modes with spectral modes, a phenomenon commonly known as mode mixing \citep{liu2011method}. The instrumental mode mixing combined with the smooth spectral nature of foregrounds results in them occupying a region in the cylindrical power spectrum known as the ``foreground wedge'' \citep{datta2010bright,vedantham2012imaging,morales2012four,liu2014epoch1,liu2014epoch2,murray2018effect}. The two main approaches used in the mitigation of foregrounds are: ``subtraction'', in which the smooth spectral nature of the foregrounds is used to model and subtract them from the visibilities, and ``avoidance'', in which the power spectrum is constructed only from the $k$ modes that fall outside the foreground wedge and are thus assumed largely free of foregrounds.

The maximum extent of the foregrounds in the cylindrical power spectrum is described by a straight line in the ($k_{\perp},k_{\parallel}$) space, commonly known as the horizon line \citep{vedantham2012imaging,morales2012four,trott2012impact,thyagarajan2013study}. However, for cylindrical power spectra estimated by phase-tracking instruments, as opposed to drift-scan approaches, the equations describing such a line are derived under the assumption that sky sources lie on the tangent plane of the sky at the phase center. This assumption leads to the complex visibility function sampled by an interferometer for an arbitrary phase center taking the form of a two-dimensional Fourier transform. As a result, the conventional horizon line equations are only valid for instruments with a small field of view (FoV) without primary beam sidelobes, where such a flat-sky approximation holds. For wide-field instruments, sky sources far away from the phase center can strongly impact the power spectrum, which these equations do not describe well. 

In this paper, we derive the horizon line equation in the cylindrical power spectrum for a pointing observation without taking a flat-sky approximation. These equations describe the actual maximum extent of foregrounds in cylindrical power spectra estimated using wide-field instruments. This is important because it helps optimize our instruments, observations, and processing for signal extraction. It is also crucial for diagnoses, by helping trace the source of power observed in the power spectrum to the emitter. We find that the predictions of these equations differ significantly from those of the conventional horizon line equations. In addition to the horizon line equation, we also derive an equation that describes the extent of contamination due to a specific source in the sky on the cylindrical power spectrum, also without a flat-sky approximation. Again, we find that the equation deviates from what we obtain from an equation derived using the flat-sky approximation. We confirm the validity of both sets of equations through realistic simulations for a range of observation parameters. We present a Python library \texttt{pslines}, that can be used to generate and plot the horizon and source lines derived in this analysis.

The paper is organized as follows: Section~\ref{sec:background} introduces the problem and illustrates the impact of pointing observations on signal delays. In Sect.~\ref{sec:max_delays}, we derive analytical expressions for the maximum delays in the visibilities. Section~\ref{sec:horizon_ps} uses these expressions to derive equations for the horizon line in the cylindrical power spectrum without imposing a flat-sky approximation. In Sect.~\ref{sec:signature_source}, we derive an expression for the modes in the cylindrical power spectrum most strongly affected by a specific point source. Section~\ref{sec:simulations} describes the simulations performed to validate the horizon and source line equations. In Sect.~\ref{sec:discussion}, we discuss the limitations of the derived equations and their potential applications in 21-cm cosmology. Section~\ref{sec:summary} summarizes the main results of this paper.

\section{Pointing observations and delays}\label{sec:background}
Power spectrum analyses performed in 21-cm cosmology can be divided into two major groups: the delay spectrum approach and the reconstructed power spectrum approach \citep{morales2019understanding}. 

In the delay spectrum approach \citep{parsons2012per}, the individual visibilities from pairs of interferometric elements are directly Fourier transformed along the frequency axis to obtain the visibilities as a function of the signal delay ($\eta$) between the pair of elements. The corresponding power spectrum is closely related to the cosmological power spectrum and contains all statistical information about the 21-cm field under the assumption of Gaussianity. This method is usually employed for drift scan telescopes where the data is not phased to a direction fixed to the sky. In this case, the maximum delay ($|\eta|^{\mathrm{max}}$) is given by:
\begin{equation}\label{eq:delay_ps}
|\eta|^{\mathrm{max}} = |\vec{b}|/c,
\end{equation}
where $\vec{b}$ is a baseline vector in units of meters, with the maximum delay corresponding to a source at the physical horizon. Therefore Eq.~(\ref{eq:delay_ps}) represents the horizon line equation for the delay power spectrum for a drift scan observation.

In the reconstructed or sky-plane power spectrum approach, the power spectrum is constructed from the gridded visibility data in the $u\varv\nu$ space. This requires choosing a phase center fixed to a direction in the sky which then moves with respect to the zenith during an observation. The raw cross-correlation data measured by the instrument needs to be phased in the direction of the phase center. This is commonly referred to as geometric delay correction, and this results in the phase center having a phase of zero across all stations. Consider a Cartesian coordinate system with axes ABC, where the C axis points along the direction of the phase center. Let $\vec{\hat{s}} = (l,m,n)$ be a unit vector in the direction of a particular source in the sky and $\vec{b}_{p} = (U_p, V_p, W_p)$ be the position vector of the $p^{\text{th}}$ interferometric element. The electric field due to a unit-flux unpolarized source at the $p^{\text{th}}$ element and frequency $\nu$ is:
\begin{equation*}
    e_{p}(\nu) = e^{-2\pi i \left(\vec{b}_{p}\cdot\hat{\vec{s}}\right)\nu/c}.
\end{equation*}
The cross-correlated raw data from the $p^{\text{th}}$ and $q^{\text{th}}$ elements are:
\begin{align*}
    V_{pq}(\nu) = e_{p}(\nu)e^{*}_{q}(\nu) &= e^{-2\pi i \left(\vec{b}_{p}-\vec{b}_{q}\right)\cdot\hat{\vec{s}}\,\nu/c}\\ 
    &= e^{-2\pi i\left(U_{pq}l + V_{pq}m + W_{pq}n\right)\nu/c},
\end{align*}
where $\vec{b}_{pq} = \vec{b}_{p}-\vec{b}_{q}$ is the baseline vector, having the components ($U_{pq}, V_{pq}, W_{pq}$) in this coordinate system. The phase center $\vec{\hat{p}}$ has the components $(0,0,1)$ and the phase in the direction of the phase center is $e^{-2\pi i W_{pq}\nu/c}$. Phasing the data amounts to applying the inverse of this phase to the data, so that the phase center has a phase of zero for all baselines. After phasing, the visibility data is:
\begin{equation}\label{eq:vcz}
    V_{pq}(\nu) = e^{-2\pi i \left(\vec{b}_{pq}\cdot(\hat{\vec{s}}-\hat{\vec{p}})\right)\nu/c} = e^{-2\pi i\left(U_{pq}l + V_{pq}m + W_{pq}(n-1)\right)\nu/c}.
\end{equation}
This equation gives the visibility at a baseline $\vec{b}_{pq}$ due to a unit-flux source $\vec{\hat{s}}$ for a phase center $\vec{\hat{p}}$, assuming noiseless isotropic receivers with unit gains.

We will now examine how phasing the data to a certain phase center affects the horizon line in the delay power spectrum. We hereby drop the $pq$ suffix. Following Eq.~(\ref{eq:vcz}) for a unit-flux unpolarized source at $\hat{\vec{s}}$, the delay spectrum is given by:
\begin{align}\label{eq:delay_expr}
V(\eta) &= \int_{\nu} e^{-2\pi i\left(Ul + Vm + W(n-1)\right)\nu/c}  e^{2\pi i \eta\nu} d\nu \nonumber\\
&= \delta(\eta-\eta_{0}),\nonumber\\
\text{where }\eta_{0} &= \vec{b}\cdot(\hat{\vec{s}}-\hat{\vec{p}})/c = \left(Ul + Vm + W(n-1)\right)/c.
\end{align}
Here $\delta(\eta-\eta_{0})$ is a Dirac delta function\footnote{Usually the Fourier transform is carried out over a finite bandwidth with a window function over frequency. Then instead of a delta function, we will have a sinc function with a width proportional to the inverse of the bandwidth, convolved with the Fourier transform of the frequency window.} with a peak at $\eta=\eta_{0}$.
Because $\hat{\vec{s}}$ is a unit vector,
\begin{align}
&-\dfrac{|\vec{b}|}{c} \leq \dfrac{\vec{b}\cdot\hat{\vec{s}}}{c} \leq \dfrac{|\vec{b}|}{c} \nonumber \\
\text{or, } &\dfrac{-|\vec{b}|-W}{c} \leq \dfrac{Ul + Vm + W(n-1)}{c} \leq \dfrac{|\vec{b}|-W}{c} \nonumber \\
\text{or, } &\dfrac{-|\vec{b}|-W}{c} \leq \eta_{0} \leq \dfrac{|\vec{b}|-W}{c}.
\label{eq:delay_ps_phased}
\end{align}
Thus, the maximum (horizon) delay is not simply $|\vec{b}|/c$. It depends on the $W$ coordinate for the particular baseline. The range of delays remains the same but gets shifted through geometric phasing due to the artificial injection of an additional delay.

\begin{figure}
    \includegraphics[width=\hsize]{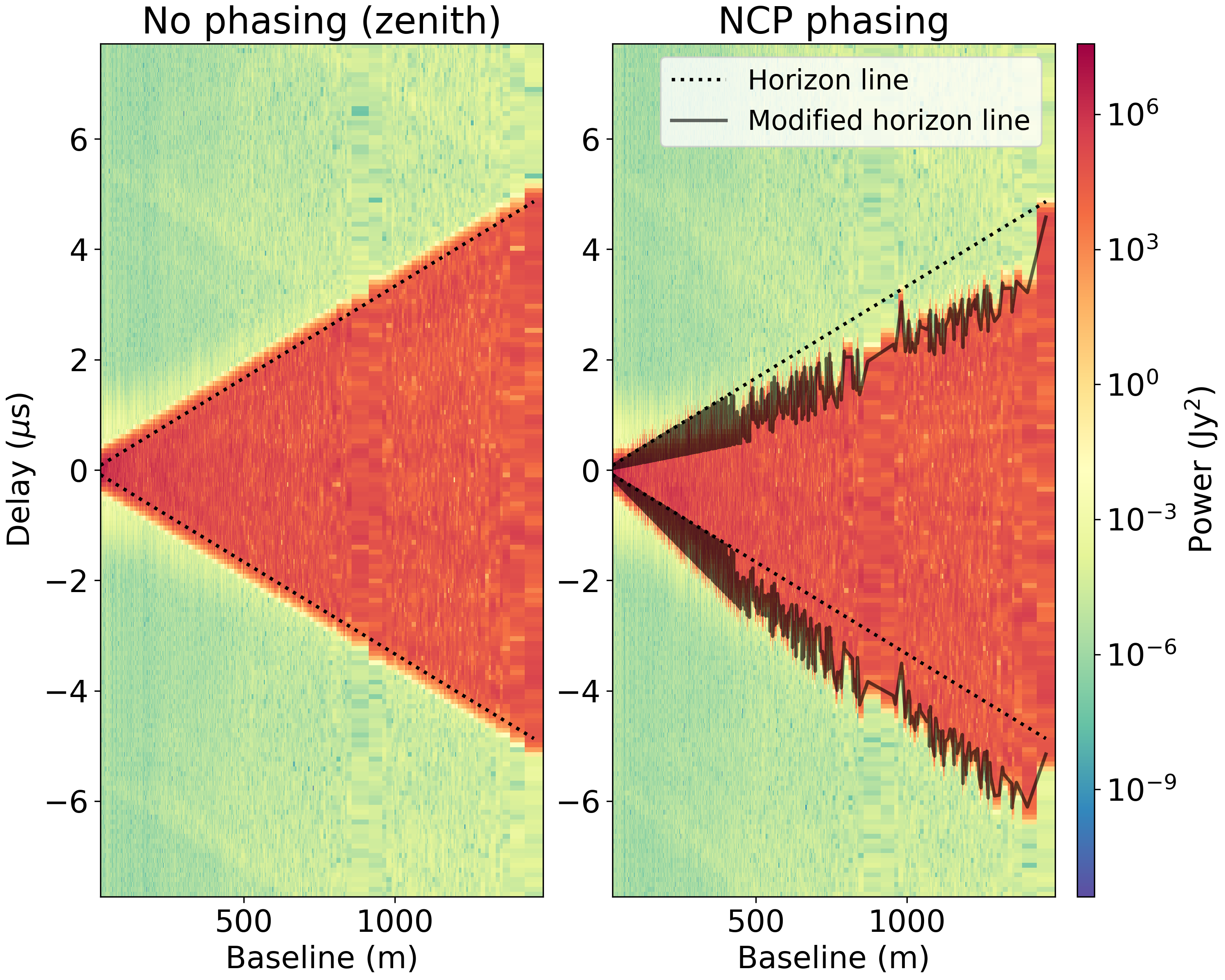}
    \caption{Effect of geometric phasing on the delay power spectrum. The two panels show the delay power spectra before (left) and after (right) phasing to the NCP for a full-sky simulation with NenuFAR. The dotted lines correspond to the standard horizon limit of Eq.~(\ref{eq:delay_ps}), and the solid lines correspond to the extended horizon due to phasing from Eq.~(\ref{eq:delay_ps_phased}).}
    \label{fig:delay_ps}
\end{figure}
To illustrate the impact of phasing the data on the delays, we constructed the delay power spectrum for visibilities from a $1\,$h full-sky simulation of a NenuFAR observation of the north celestial pole (NCP). The details of the sky model and simulation parameters are described in Sect.~\ref{sec:simulations}. In Fig.~\ref{fig:delay_ps}, we show the delay power spectra before and after phasing the data to the NCP. While plotting, the delay spectra of all baselines with a negative $W$ component have been flipped along the delay axis. This effectively makes sure that all baselines have a positive component toward the north (i.e. the vector joining the $p^{\text{th}}$ element to the $q^{\text{th}}$ element has a positive component toward the north). As a result, before phasing, positive delays correspond to sources toward the north of the zenith and negative delays correspond to sources toward the south of the zenith. The dotted lines correspond to Eq.~(\ref{eq:delay_ps}), while the solid lines indicate the horizon line from Eq.~(\ref{eq:delay_ps_phased}). We see that the two equations correctly describe the horizon limit, before and after phasing the data. There is a slight spillover due to spectral leakage coming from the Blackman-Harris filter which was applied to the data along frequency before the Fourier transformation. This shift in the delays due to phasing can be intuitively understood as a shift of the coordinate system such that the phase center has a delay of zero. Since the NCP is toward the north of the zenith, phasing the data to the NCP decreases the horizon toward the north of the phase center, and increases the horizon limit toward its south. However, it should be noted that this is not a simple rotation of the coordinate system, and the delay due to sources $\pi/2$ away from the phase center is no longer confined to $|\vec{b}|/c$ after phasing (discussed in Appendix \ref{sec:ncp_phasing}). Normally we plot only the absolute values of the delay, and in that case, the negative delays will spill over the traditional horizon line.

It should be noted that this plot is possible for a long observation (not a snapshot) only when the phase center is the NCP or the south celestial pole (SCP) where the $W$ coordinates are independent of time. Additionally, the lines can be plotted as long as the data is not binned in the baseline direction, as the $W$ values are dependent on the baseline directions. This simple example illustrates how the classic horizon delay lines are modified for interferometers that phase-track a fixed point in the sky. In the next section, we derive a more complete description of maximum delays for phase tracking instruments.

\section{Maximum Delays}\label{sec:max_delays}
Evaluating the maximum delay from Eq.~(\ref{eq:delay_ps_phased}) requires the knowledge of the individual $W$ coordinates for the baseline under consideration. Thus, Eq.~(\ref{eq:delay_ps_phased}) cannot describe a situation where multiple baselines oriented in different directions are averaged together, as is the case for the cylindrical power spectrum. Additionally, for any phase center other than the NCP and SCP, the $W$ coordinates, and consequently the horizon lines, will shift with time. In an observation, we will typically have a large number of baselines oriented in different directions and a large number of sources distributed throughout the sky. In this section, we derive general equations for the horizon delay as a function of baseline length similar to Eq.~(\ref{eq:delay_ps}), but accounting for the phasing to the phase center.

\begin{figure}
    \includegraphics[width=\hsize]{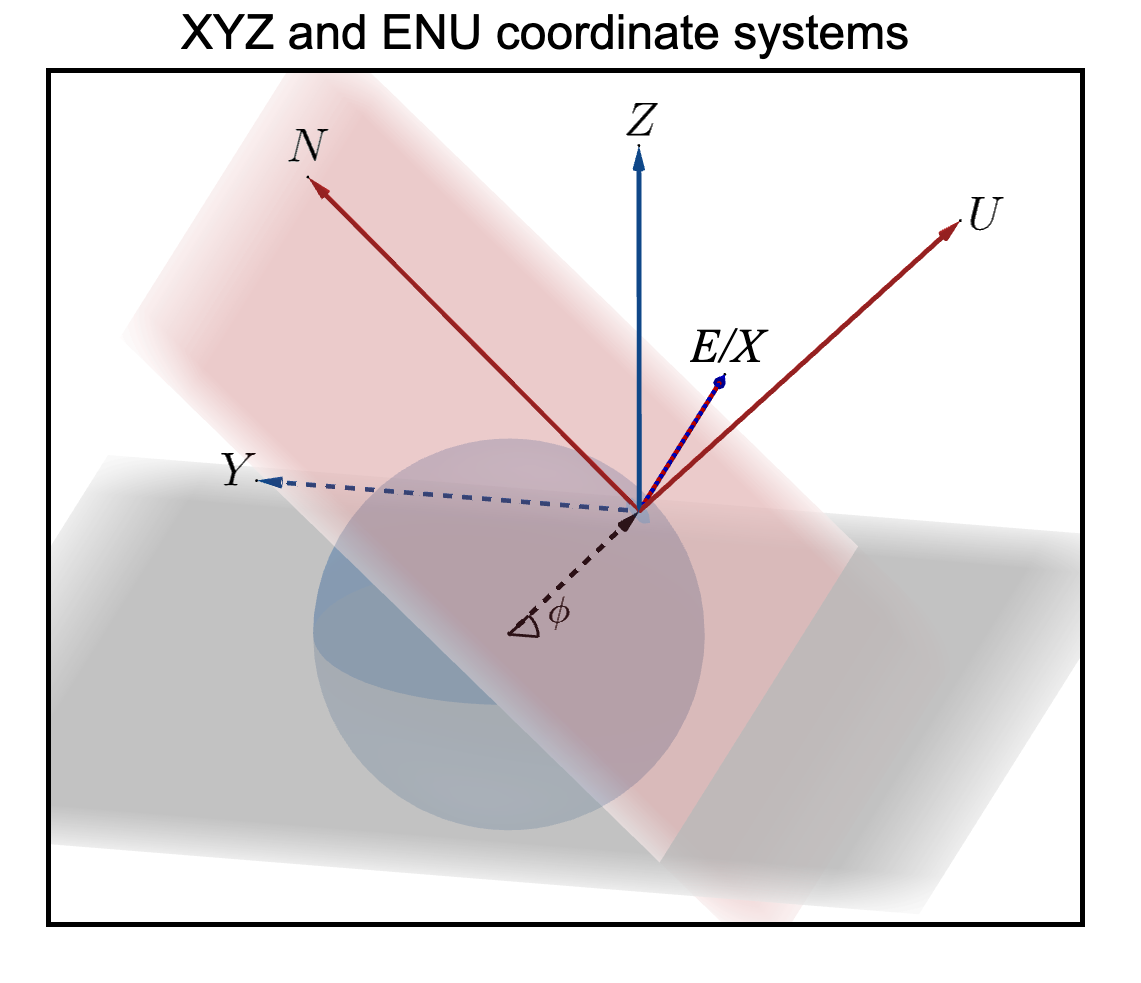}
    \caption{The coordinate systems used in the calculations. The origin of the coordinate system is placed at the location of the array for illustration. Since the sources are at an infinite distance, the exact location of the coordinate center is irrelevant.}
    \label{fig:coord}
\end{figure}

\subsection{Analytical Treatment}\label{sec:analytical}
Here, we assume a three-dimensional coordinate system XYZ fixed to the Earth with the Z axis pointing toward the NCP, the X axis pointing due east, and the Y axis pointing in the direction Z$\times$X. Consider another three-dimensional coordinate system ENU (east-north-up) where U points to the zenith, N points to the north along the plane perpendicular to the zenith, and E points to the east. The two coordinate systems are illustrated in Fig.~\ref{fig:coord}. $H, \delta$ and $H_0, \delta_0$ are the hour angle, declination (Dec) of the source ($\hat{\vec{s}}$) and the phase center ($\hat{\vec{p}}$) respectively, and $\phi$ is the latitude of the location of the telescope. Table \ref{tab:notation} summarizes the notation used in the paper. The coordinates of the baseline ($\vec{b}$), source ($\hat{\vec{s}}$), zenith ($\hat{\vec{z}}$) and the phase center ($\hat{\vec{p}}$) in these systems are:
\begin{align*}
\vec{b} &= (B_E, B_N, B_U) \text{ [ENU]}\\
&= (B_E, B_N \sin\phi-B_U \cos\phi, B_N \cos\phi + B_U\sin\phi) \text{ [XYZ]},\\
\hat{\vec{s}} &= (-\cos\delta\sin H, -\cos\delta\cos H, \sin\delta) \text{ [XYZ]},\\
\hat{\vec{z}} &= (0, -\cos\phi, \sin\phi) \text{ [XYZ]\footnotemark},\\
\hat{\vec{p}} &= (-\cos\delta_0\sin H_0, -\cos\delta_0\cos H_0, \sin\delta_0) \text{ [XYZ]}.
\end{align*}\footnotetext{It should be noted that $\hat{\vec{z}}$ points along the U axis, not along the Z axis.}We now assume that all baselines lie on the plane perpendicular to the zenith (or the Earth's tangent plane, considering the Earth to be a sphere), so $B_U=0$. This is a fair assumption for EoR or CD experiments where the baselines used to construct the power spectrum are small relative to the curvature of the Earth. The components can now be written in terms of the baseline length $|\vec{b}|$ and the angle $\theta$ it makes with the E axis. In this analysis we assume that the array has baselines in all directions and $\theta$ can take any value between 0 and $2\pi$. This is a fair assumption for imaging interferometers with dense cores used in 21-cm cosmology. The coordinates of the baseline are then given by:
\begin{equation*}
\vec{b} = |\vec{b}|\,(\cos\theta, \sin\theta \sin\phi, \sin\theta \cos\phi) \text{ [XYZ]}.
\end{equation*}
The corresponding expression for the delay due to a source ($\eta_0$ from Eq. (\ref{eq:delay_expr})) is given by:
\begin{align}\label{eq:delay_arbitrary}
\eta_0=&\dfrac{|\vec{b}|}{c}\, \Bigl(\cos\theta\cos\delta_0\sin H_0-\cos\theta\cos\delta\sin H\nonumber\\&+\sin\theta\sin\phi\cos\delta_0\cos H_0-\sin\theta\sin\phi\cos\delta\cos H\nonumber\\&+\sin\theta\cos\phi\sin\delta-\sin\theta\cos\phi\sin\delta_0\Bigr)
\end{align}
The condition that both $\hat{\vec{s}}$ and $\hat{\vec{p}}$ are above the physical horizon takes the form:
\begin{align}\label{eq:horizon_mask_arbitrary}
\hat{\vec{s}}\cdot\hat{\vec{z}} \geq 0 &\text{ or, } \cos\delta\cos H\cos\phi + \sin\delta\sin\phi \geq 0;\nonumber\\
\hat{\vec{p}}\cdot\hat{\vec{z}} \geq 0 &\text{ or, }\cos\delta_0\cos H_0\cos\phi + \sin\delta_0\sin\phi \geq 0.
\end{align}
\begin{table}[]
\caption{The notation used throughout this paper.}
\label{tab:notation}
\begin{tabular}{@{}ll@{}}
\toprule
Notation                  & Explanation                                              \\ \midrule\midrule
$\vec{b}$                  & Baseline vector (in meter)                               \\
$\vec{b'}$                  & Baseline vector projected on the $u\varv$ plane (in meter)         \\
$\vec{u}$                  & Baseline vector projected on the $u\varv$ plane (in $\lambda$)     \\
$\hat{\vec{s}}$            & Source (unit) vector                                     \\
$\vec{s'}$                 & Source vector projected on the tangent plane to the\\& sky at the phase center            \\
$\hat{\vec{p}}$            & Phase center (unit) vector                               \\
$\hat{\vec{z}}$            & Zenith (unit) vector     
                    \\
$\vec{a}$            & Projection of $\hat{\vec{s}}-\hat{\vec{p}}$ on the plane perpendicular to $\hat{\vec{z}}$     
                    \\ \midrule
ABC                       & Coordinate system with C pointing along \\& the direction of phase center   \\
XYZ                       & Coordinate system with Z pointing along \\& the Earth's axis \\
ENU                       & East-north-up coordinate system                          \\ \midrule
$H, \alpha, \delta$       & Hour angle, RA, Dec of source                            \\
$H_0, \alpha_0, \delta_0$ & Hour angle, RA, Dec of phase center                      \\
$t$                       & Local siderial time                                      \\
$\phi$                    & Latitude                                                 \\
$\theta$                  & Angle between a baseline on the Earth's tangent\\& plane and the east direction        \\
$a_0$                     & Altitude (elevation) of the phase center                                 \\
$\psi$                    & Angle between the source and the phase center          \\
$\beta$                    & Angle between $\vec{u}$ and $\vec{s'}$
\\ \bottomrule
\end{tabular}
\end{table}We would like to obtain the extrema of Eq.~(\ref{eq:delay_arbitrary}) for a given $\delta_0$, subject to the horizon condition (Eq.~(\ref{eq:horizon_mask_arbitrary})). Instead of solving the full three-dimensional equation $\nabla \eta_0(H, \delta, H_0, \theta) = 0$, we first reduce the problem to the plane containing the zenith vector and the Earth's axis (the meridian plane). It should be noted that for any phase center other than the NCP and SCP, this reduction of the problem cannot be done without loss of generality, since $H_0$ affects the delay values and the extrema do not necessarily lie on the meridian plane. We will show in Sect. \ref{sec:full_synthesis} that this reduction is valid only in the regime of $|\delta_0 +\phi| >\pi/2$. The situation for the NCP is described in Appendix~\ref{sec:ncp_phasing}. Nevertheless, to gain insight into the problem, we investigate the situation of a north-south baseline with the phase center and source both lying on the meridian plane. To get an extreme case of maximum delay, we choose $H_0$ and $H$ so that the phase center and the source can be located the furthest apart. This amounts to setting $H=0$, $H_0=\pi$, and $\theta=\pi/2$ in Eq.~(\ref{eq:delay_arbitrary}). The equation simplifies to:
\begin{align*}
\eta_0(\phi, \delta, \delta_0) = \dfrac{|\vec{b}|}{c}\, &\Bigl(-\sin\phi \cos\delta_0 -\sin\phi \cos\delta + \cos\phi \sin\delta \\&- \cos\phi\sin\delta_0\Bigr).
\end{align*}
The declination of the source responsible for the maximum absolute delay is analytically obtained by setting $\frac{\partial \eta_0}{\partial \delta} = 0$. This gives us:
\begin{align}\label{eq:horizon_arbitrary_partial}
\delta\Bigr|_{\mathrm{max}} &= \begin{cases} \phi - \pi/2 & \mbox{ if } \phi \geq 0\\ \phi + \pi/2 & \mbox{ if } \phi \leq 0
\end{cases}\nonumber\\
\text{and }|\eta_0|^{\mathrm{max}} &= \begin{cases} |\vec{b}|\,\Bigl(1+\sin\bigl(\delta_0 + \phi\bigr)\Bigr)/c & \mbox{ if } \phi \geq 0\\ |\vec{b}|\,\Bigl(1-\sin\bigl(\delta_0 + \phi\bigr)\Bigr)/c & \mbox{ if } \phi \leq 0.
    \end{cases}
\end{align}
The maximum delays occur due to a source at the physical horizon\footnote{For an array located at the equator ($\phi=0$), there are two situations where the source is on the physical horizon, for $\delta=\pm \pi/2$. We note, however, that for $\phi=0$, Eq. (\ref{eq:horizon_arbitrary_partial}) is never valid since the regime $|\delta_0 + \phi| > \pi/2$ is impossible. This is evident from Eq. (\ref{eq:horizon_arbitrary}).}, hence it is not necessary to perform the maximization subject to Eq.~(\ref{eq:horizon_mask_arbitrary}). These equations reduce to the ones obtained for NCP phasing at $\delta_0 = \pi/2$ and $\phi \geq 0$ (Eq.~(\ref{eq:horizon_ncp}) in Appendix~\ref{sec:ncp_phasing}). In the next section, we show that this equation is only valid for a subset of all possible situations since the assumption that the maximum delays will occur when the source and phase center lie on the meridian plane is not always correct.

\subsection{Full Synthesis}\label{sec:full_synthesis}
We consider a full synthesis when $H_0$ takes all values between 0 and $2\pi$ (i.e., a $24\,$h synthesis) during the observation due to the rotation of the Earth. At each such time, there can be a source anywhere in the sky and any possible baseline can be responsible for the maximum delay. The expression we want to maximize is $\vec{b}\cdot(\hat{\vec{s}}-\hat{\vec{p}})/c$. Now $|\hat{\vec{s}}-\hat{\vec{p}}|$ is maximum when $\hat{\vec{s}}$ and $\hat{\vec{p}}$ point in the opposite directions, and in that case $(\hat{\vec{s}}-\hat{\vec{p}})$ has a magnitude of 2. Additionally, $\vec{b}\cdot(\hat{\vec{s}}-\hat{\vec{p}})$ is maximum when $\vec{b}$ is in the same direction as $(\hat{\vec{s}}-\hat{\vec{p}})$. Let us only consider the situation when the phase center is above the horizon for at least a part of the full synthesis, where $|\delta_0-\phi| \leq \pi/2$. Now we consider two cases:
\begin{itemize}
\item When $|\delta_0+\phi| \leq \pi/2$ (i.e., when the phase center is non-circumpolar), at some point during a full synthesis, $\hat{\vec{p}}$ will go below the horizon. Whenever $H_0$ assumes a value such that the phase center crosses the horizon, $H$ and $\delta$ get optimized to make $\hat{\vec{s}}$ point along the opposite direction of $\hat{\vec{p}}$. Finally $\theta$ is optimized to have $\vec{b}$ pointing along $\hat{\vec{p}}$ or $-\hat{\vec{p}}$. Since the baselines are constrained to lie on the Earth's tangent plane, this situation can occur when $\hat{\vec{s}}$, $\hat{\vec{p}}$ and $\vec{b}$ all lie on the physical horizon, irrespective of whether we impose the horizon condition of Eq.~(\ref{eq:horizon_mask_arbitrary}). Thus, the maximum delay is always $2 |\vec{b}|/c$. 
\item When $|\delta_0+\phi| > \pi/2$ (i.e., when the phase center is circumpolar), $\hat{\vec{p}}$ will always be above the horizon and can never cross it. Here the maximum delay occurs when $\hat{\vec{p}}$ is closest to the horizon, which by definition lies on the meridian plane. Thus, $H_0=\pi$ and the assumption taken to derive Eq.~(\ref{eq:horizon_arbitrary_partial}) is satisfied and Eq.~(\ref{eq:horizon_arbitrary_partial}) is valid in this regime.
\end{itemize}
In order to account for both regimes of $|\delta_0+\phi| \leq \pi/2$ and $|\delta_0+\phi| > \pi/2$, Eq.~(\ref{eq:horizon_arbitrary_partial}) is modified to have the form:
\begin{align}\label{eq:horizon_arbitrary}
|\eta_0|^{\mathrm{max}} = \dfrac{|\vec{b}|}{c}\biggl(1+\sin\bigl(\max\bigl\{|\phi+\delta_0|, \pi/2\bigr\}\bigr)\biggr) \text{    for } |\delta_0-\phi| \leq \pi/2.
\end{align}

\begin{figure}
    \includegraphics[width=\hsize]{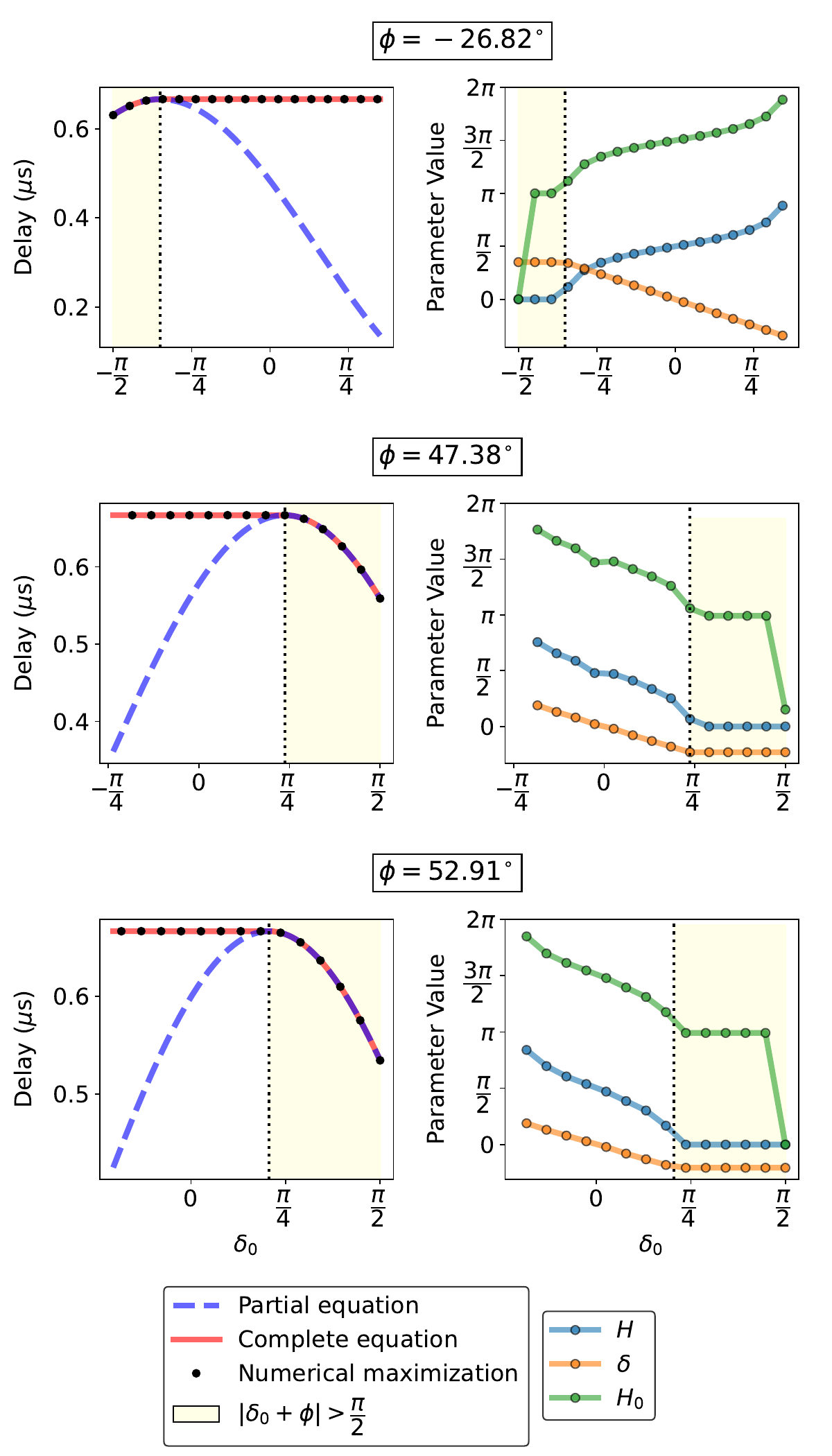}
    \caption{The maximum delay in visibilities measured by a $100\,$m baseline, for a given phase center declination, assuming a full synthesis. Left: The maximum delays as a function of declination of the phase center $\delta_0$, obtained by numerical maximization of Eq.~(\ref{eq:delay_arbitrary}) over $H, \delta, H_0, \theta$ (black dots), the maximum delays predicted by the partial equation (Eq. (\ref{eq:horizon_arbitrary_partial}): blue dashed line) and those predicted by the full equation (Eq. (\ref{eq:horizon_arbitrary}): red solid line). Right: The values of $H$, $\delta$, and $H_0$ at the maximum. The vertical dotted lines separate regimes where $|\delta_0+\phi|\leq\pi/2$ and $|\delta_0+\phi|>\pi/2$.}
    \label{fig:max_latitudes}
\end{figure}

We verify Eq. (\ref{eq:horizon_arbitrary}) by numerically maximizing Eq.~(\ref{eq:delay_arbitrary}) over $H, \delta, H_0$, and $\theta$, subject to Eq.~(\ref{eq:horizon_mask_arbitrary}) for a given $\phi$ and $\delta_0$. This was done for a baseline length of $100\,$m, by evaluating Eq.~(\ref{eq:delay_arbitrary}) on a discrete grid in $H, \delta, H_0, \theta$ space, masking out regions where Eq.~(\ref{eq:horizon_mask_arbitrary}) is not valid and finding the extrema. To decrease numerical errors due to finite grid size at some $\delta_0$ values, the optimized parameter values were used as initial values for the Nelder-Mead simplex algorithm \citep{gao2012implementing} to obtain the final extrema. We repeated this process for a series of phase center declinations ($\delta_0$), for the latitudes $\phi = -26.82^{\circ}$, $47.38^{\circ}$, and $52.91^{\circ}$ (corresponding to SKA, NenuFAR, and LOFAR respectively). In the left panels of Fig.~\ref{fig:max_latitudes}, we see that Eq.~(\ref{eq:horizon_arbitrary_partial}) (blue dashed line) agrees with the numerically maximized delays only for $|\delta_0+\phi|>\pi/2$. For all other values of $\delta_0$, the numerically maximized delay is $2 |\vec{b}|/c$. Eq.~(\ref{eq:horizon_arbitrary}) (red solid line) agrees with the numerically maximized delays in both regimes $|\delta_0+\phi|>\pi/2$ and $|\delta_0+\phi|\leq\pi/2$. It should be noted that even if Eq.~(\ref{eq:horizon_mask_arbitrary}) is not imposed, we get the same result, thus confirming the claim that the maximization results should be unaffected by whether or not we impose the condition that the source and phase center lie above the physical horizon.

In the right panels of Fig.~\ref{fig:max_latitudes}, we see that in the regime $|\delta_0+\phi| \leq \pi/2$, $H$ and $H_0$ are separated by $\pi$ and the plot of $\delta$ against $\delta_0$ has a slope of $-1$. This shows that the source vector and the phase center vectors are in diametrically opposite directions as explained previously. Also, in this regime, $H$ and $H_0$ will give the same delay for two values separated by $\pi$. These $\pi$ jumps between consecutive points have been removed while plotting for visual clarity, along with any $2\pi$ jumps. In the regime, $|\delta_0+\phi| > \pi/2$, the source and phase center are no longer in opposite directions, and $\delta|_{\mathrm{max}}$ agrees with the values predicted by Eq.~(\ref{eq:horizon_arbitrary_partial}). Also, $H_0$ assumes a value of $\pi$ (except at $\delta_0=\pi/2$ or $-\pi/2$ where the value of $H_0$ is irrelevant). Again, this confirms that the assumption made in deriving Eq.~(\ref{eq:horizon_arbitrary_partial}) is correct in this regime.

\subsection{Snapshot at a given LST}\label{sec:given_lst}
In the previous section, we investigated the situation of a full synthesis where the phase center moves with respect to the array and derived an expression for the maximum delay over this $24\,$h period. In this section, we instead consider a snapshot observation at a given LST, where the phase center is fixed with respect to the array. The previous analysis shows that in both regimes ($|\delta_0+\phi| \leq \pi/2$ and $|\delta_0+\phi| > \pi/2$), the maximum delay occurs due to a source on the physical horizon, lying on the plane containing $\hat{\vec{z}}$ and $\hat{\vec{p}}$, and in the opposite direction of the phase center. Using this insight, we can derive a general equation for the horizon delay for an arbitrary $H_0$. At a given local sidereal time (LST), the maximum additional delay due to phasing is equal to the $W$ coordinate of a baseline lying in this plane. This additional delay is $\vec{b}\cdot\hat{\vec{p}}=|\vec{b}|\cos a_0$, where $a_0$ is the altitude of the phase center. Thus, the equation for the maximum delay is given by:
\begin{equation*}
|\eta_0|^{\mathrm{max}} = \dfrac{|\vec{b}|}{c}\,\Bigl(1+\cos a_0\Bigr).
\end{equation*}
For a given LST (denoted by $t$) and right ascension (RA) of the phase center ($\alpha_0$), this reduces to:
\begin{align}\label{eq:horizon_arbitrary_lst}
|\eta_0|^{\mathrm{max}}=\dfrac{|\vec{b}|}{c}\left(1+\sqrt{1-\bigl(\sin\phi\sin\delta_0 + \cos\phi\cos\delta_0\cos (t-\alpha_0)\bigr)^2}\right).
\end{align}
\begin{figure}
    \includegraphics[width=\hsize]{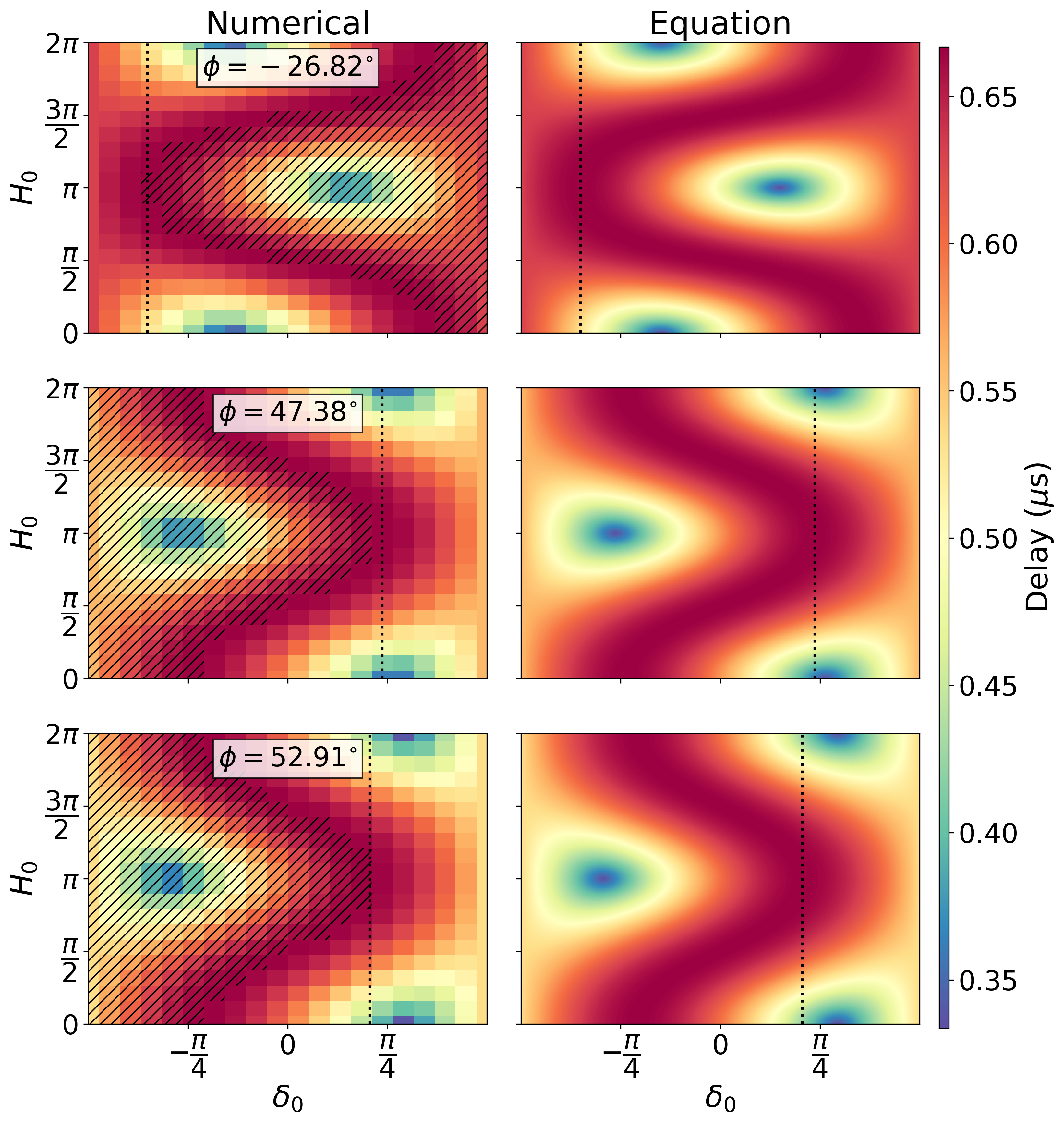}
    \caption{The maximum delay in visibilities measured by a $100\,$m baseline for a given phase center location in the sky ($H_0, \delta_0$). Left: The maximum delays, obtained by numerical maximization of Eq.~(\ref{eq:delay_arbitrary}) over $H, \delta$, and $\theta$. Right: The maximum delays expected from Eq.~(\ref{eq:horizon_arbitrary_lst}). The vertical dotted lines separate regimes where $|\delta_0+\phi|\leq\pi/2$ and $|\delta_0+\phi|>\pi/2$. The slanted stripes indicate the regions where the phase center is below the physical horizon.}
    \label{fig:max_latitudes_hadec}
\end{figure}We verify this equation by maximizing Eq.~(\ref{eq:delay_arbitrary}) over $H, \delta$, and $\theta$ at a particular $H_0, \delta_0$, both with and without imposing the horizon condition of Eq.~(\ref{eq:horizon_mask_arbitrary}). We repeated this for multiple latitudes to verify Eq.~(\ref{eq:horizon_arbitrary_lst}) for a range of combinations of $\phi, H_0$, and $\delta_0$. Figure~\ref{fig:max_latitudes_hadec} shows the maximum delay as a function of the phase center direction and latitude. The left column shows the values obtained by maximizing Eq.~(\ref{eq:delay_arbitrary}) at each $H_0, \delta_0$. The right column shows the expected maximum delays obtained using Eq.~(\ref{eq:horizon_arbitrary_lst}). We find that the equation can predict the maximum delays, which confirms the result of Sect.~\ref{sec:full_synthesis} and the premise in deriving Eq.~(\ref{eq:horizon_arbitrary_lst}) that the maximum occurs due to a source on the physical horizon, lying on the plane containing zenith and the phase center and in the direction opposite to the phase center. Also, in the regime of $|\delta_0+\phi| \leq \pi/2$, the maximum values of the delay occur along the edge of the striped regions, which indicate the physical horizon. Thus, we affirm the result from Sect.~\ref{sec:full_synthesis} that in the regime of $|\delta_0+\phi| \leq \pi/2$, the maximum delay occurs when $H_0$ and $\delta_0$ are such that the phase center lies on the horizon. In the regime of $|\delta_0+\phi| > \pi/2$, we see that the maximum value of the delay ($2 |\vec{b}|/c = 6.66\times 10^{-7}$ s for $|\vec{b}|=100\,$ m) is not reached at any $H_0$, since the phase center can never be on the horizon for this declination range. Also, the maximum occurs at $H_0=\pi$, which agrees with the result from Sect.~\ref{sec:full_synthesis} and confirms again that the assumption made in deriving Eq.~(\ref{eq:horizon_arbitrary_partial}) is valid in this regime.

We now have two equations that give us the maximum delays for a given phase center, one assuming full synthesis (Eq.~(\ref{eq:horizon_arbitrary})) and the other for a snapshot at a particular LST (Eq.~(\ref{eq:horizon_arbitrary_lst})). Both equations have been derived assuming that the sky is filled with sources, the array has baselines in all directions, and the baselines lie on the plane perpendicular to the zenith.

\section{Horizon wedge in the cylindrical power spectrum}\label{sec:horizon_ps}
In the reconstructed power spectrum approach, the visibilities are phased to the phase center direction and then gridded in the $u\varv$ plane before Fourier transformation along frequency. The gridded data is then used to construct the cylindrical power spectrum. In this section, we derive the equations governing the horizon line in the cylindrical power spectrum.

We work in the same ABC coordinate system as Sect.~\ref{sec:background} with the C axis pointing in the direction of the phase center. Rewriting Eq.~(\ref{eq:vcz}) after dropping the $pq$ suffix, we get:
\begin{equation}\label{eq:vcz_nopq}
    V(\nu) = e^{-2\pi i \vec{b}\cdot(\hat{\vec{s}}-\hat{\vec{p}})\nu/c} = e^{-2\pi i\left(Ul + Vm + W(n-1)\right)\nu/c}.
\end{equation}
Under the flat-sky approximation $l,m\ll1$. Hence $n = \sqrt{1-l^2-m^2} \approx 1$, and Eq. (\ref{eq:vcz_nopq}) becomes:
\begin{equation}\label{eq:vcz_flat}
    V(\nu) \approx e^{-2\pi i(Ul + Vm)\nu/c} = e^{-2\pi i \vec{b'}\cdot\vec{s'}\nu/c},
\end{equation}
where $\vec{b'} = (U,V)$ and $\vec{s'} = (l,m)$ are both two-dimensional vectors lying in the tangent plane of the sky perpendicular to $\hat{\vec{p}}$. Here $\vec{s'}$ is not a unit vector. For a source located at an angle $\psi$ from the phase center $\hat{\vec{p}}$, $|\vec{s'}| = \sin\psi$, and for large $\psi$, the flat-sky assumption is broken.

To construct the cylindrical power spectrum, the visibilities are placed on the $u\varv$ plane, where $u=U\nu/c$ and $\varv=V\nu/c$. In the case of an ``ideal" interferometer with full $u\varv$ coverage, for each point in the $u\varv$ plane, there is always a baseline crossing that point for each frequency channel. In this case, there will be no fluctuation in the visibility amplitude along frequency at a given $u\varv$ point, because the entire frequency dependence has been absorbed in the $u\varv$ term. Thus the entire foreground power is confined to zero delay and there is no foreground wedge. However, no interferometer is ideal, since the $u\varv$ coverage is sparse and the measurements are sampled at a discrete set of frequencies. Thus, a $u\varv$ grid with a finite-sized $u\varv$ cell is chosen and the observed visibilities are placed on this grid using a convolution kernel. Therefore, the gridded visibility at a certain $u\varv$ cell has contribution from visibilities lying within the area of support of the convolution function \citep[e.g.,][]{offringa2019precision}. So even in the sky-plane power spectrum, we end up with a horizon wedge.

Since we do not want to impose the flat-sky approximation, we must work directly with Eq.~(\ref{eq:vcz_nopq}). Let us consider a baseline for which we reach the extreme case where it stays in the same $u\varv$ cell for all frequencies\footnote{For very large FoV images or wide bandwidths, a $u\varv$ cell may not contain an entire baseline over the full bandwidth. However, in practice, gridding is performed with a convolution kernel, and a specific $u\varv$ point will still have contribution from different frequencies of the same baseline. So the same mode mixing effect will occur.}. This extreme case corresponds to the situation of the maximum delay due to mode mixing, whereas if multiple baselines cross the $u\varv$ cell at different frequencies, the mode-mixing effect would be lower. Suppose the baseline crosses the point $u,\varv$ at the central frequency $\nu_0$ ($u = U\nu_0/c$, $\varv = V\nu_0/c$). This visibility in such a $u\varv$ cell is now Fourier transformed along frequency:
\begin{align}\label{eq:delay_source_noflat}
V(\eta) &= \int_{\nu} e^{-2\pi i \vec{b}\cdot(\hat{\vec{s}}-\hat{\vec{p}})\nu/c} e^{2\pi i \eta\nu} d\nu\nonumber\\
&= \delta(\eta-\eta_{0}) \text{, where }\eta_{0} = \vec{b}\cdot(\hat{\vec{s}}-\hat{\vec{p}})/c.
\end{align}
We note that $\eta_0$ here has the same expression as the signal delay after phasing from Eq. (\ref{eq:delay_expr}). In Sect.~\ref{sec:analytical}, we have derived the maximum values of $\eta_0$ for two cases: full synthesis in Sect.~\ref{sec:full_synthesis} and at a given LST in Sect.~\ref{sec:given_lst}. We consider these two cases separately for the horizon line equation.

\subsection{Snapshot at a given LST}
We first consider the case of a particular LST. Inserting the expression for maximum delay\footnote{In the context of the cylindrical power spectrum, whenever we mention maximum delay, we refer to the maximum absolute delay value.} from Eq.~(\ref{eq:horizon_arbitrary_lst}) in Eq.~(\ref{eq:delay_source_noflat}) gives $\eta^{\mathrm{max}}_{0} = |\vec{b}|(1+\cos a_0)/c$. Now, from Sect.~\ref{sec:given_lst}, we know that the maximum delay occurs due to a source lying on the physical horizon, in the opposite direction of the phase center, on the plane containing $\hat{\vec{z}}$ and $\hat{\vec{p}}$. The baseline which will have the maximum delay due to this source will be oriented such that it lies on this same plane. The length of the projection of such a baseline on the plane perpendicular to the phase center is given by $|\vec{b'}| = |\vec{b}| \sin a_0$ where $a_0$ is the altitude of the phase center at that LST. The maximum delay is then given by:
\begin{align*}
     \eta^{\mathrm{max}}_{0} = \dfrac{|\vec{b'}|}{c \,\sin(a_0)}\,\biggl(1+\cos a_0\biggr) =\dfrac{\sqrt{u^2+\varv^2}}{\nu_0}\,\left(\dfrac{1+\cos a_0}{\sin a_0}\right).
\end{align*}
The parameters $\sqrt{u^2+\varv^2}$ and $\eta$ are related to the cosmological coordinates as \citep{morales2004toward,mcquinn2006cosmological}:
\begin{align}\label{eq:conversion}
\sqrt{u^2+\varv^2} = k_{\perp}\dfrac{D_{M}(z)}{2\pi} \text{; }
\eta = k_{\parallel}\dfrac{c(1+z)^2}{2\pi \mathcal{H}_0 \nu_{21}E(z)},
\end{align}
where $\text{D}_{M}(z)$ is the conversion factor from angular units to comoving distance units, $\mathcal{H}_{0}$ is the Hubble constant, $\nu_{21}$ = 1420~MHz and $E(z)$ is the dimensionless Hubble parameter. Converting to cosmological coordinates, we get:
\begin{align}\label{eq:horizon_line_lst}
&k^{\mathrm{LST}}_{\parallel} = k_{\perp}\dfrac{D_M(z)\mathcal{H}_0 E(z)}{c(1+z)}\left(\dfrac{1+\cos a_0}{\sin a_0}\right),\nonumber\\
&\textrm{ where } a_0 = \sin^{-1}\bigl(\sin\phi\sin\delta_0 + \cos\phi\cos\delta_0\cos(t-\alpha_0)\bigr).
\end{align}

\subsection{Full synthesis}\label{sec:ps_full_synthesis}
For full synthesis, we do a similar treatment on Eq.~(\ref{eq:horizon_arbitrary}). The horizon delay is then:
\begin{align*}
    \eta^{\mathrm{max}}_{0} = \dfrac{|\vec{b'}|}{c \, \sin a_0}\,\biggl(1+\sin\Bigl(\max\bigl\{|\phi+\delta_0|, \pi/2\bigr\}\Bigr)\biggr).
\end{align*}
From Sect.~\ref{sec:full_synthesis}, we know that for $|\delta_0+\phi| \leq \pi/2$, the maximum delay occurs when the phase center is at the horizon. Thus, $a_0 = 0$ and the horizon line has a slope of infinity. For $|\delta_0+\phi| > \pi/2$, the maximum delay occurs when the phase center crosses the meridian plane closest to the horizon ($H_0=\pi$). Thus, $\sin a_0 = \sin\phi\sin\delta_0 + \cos\phi\cos\delta_0\cos H_0 = -\cos(\delta_0+\phi)$. Inserting these into the expression for maximum delay and converting to cosmological coordinates using Eq.~(\ref{eq:conversion}), the horizon line equation takes the form:
\begin{align}\label{eq:horizon_line_full}
{\small
    k^{\mathrm{syn}}_{\parallel}= \begin{cases} k_{\perp}\dfrac{D_M(z)\mathcal{H}_0 E(z)}{c(1+z)}\,\left( \dfrac{1+\sin(|\delta_0+\phi|)}{-\cos(\delta_0+\phi)}\right) & \mbox{if } |\delta_0+\phi| > \pi/2 \\\\
    \hspace{2cm}\mathrm{Undefined} & \mbox{if } |\delta_0+\phi| \leq \pi/2. \end{cases}
}
\end{align}
It should be noted that a full synthesis observation is seldom performed over a single day. However, for target fields close to the celestial poles, and for deep integrations spanning several hundred hours over multiple months, it is likely that the full $24\,$h of LST will be covered. The power spectrum estimated from the combined data will have foreground power reaching up to the full synthesis line. Equation (\ref{eq:horizon_line_full}) thus indicates the maximum possible foreground wedge extent for a chosen phase center, without considering additional details of observation parameters.

We now have two equations: Eq.~(\ref{eq:horizon_line_lst}) and Eq.~(\ref{eq:horizon_line_full}), that describe the horizon line in the cylindrical power spectrum and have been derived without imposing a flat-sky approximation. It should be noted that the cylindrical power spectrum constructed in Cartesian coordinates is not the correct basis for describing curved sky effects, and this can be achieved by using a spherical harmonic spatial basis. However, if the cylindrical power spectrum is still used, these derived equations correctly describe the extent of full-sky foregrounds in it. This is the case for most tracking interferometers where the main lobe of the primary beam is relatively narrow and the 21-cm signal of interest is dominated by its footprint, while the bright foregrounds picked up by the primary beam sidelobes can still contaminate the power spectrum up to the horizon lines given by these equations. In Sect.~\ref{sec:sim_horizon}, we use full-sky simulations to validate both equations.

\subsection{Special case of zenith phasing}\label{sec:drift_scan}
Starting with Eq. (\ref{eq:vcz_flat}) which assumes a flat-sky approximation and considering an extreme case of a baseline that remains in the same $u\varv$ cell for all frequency channels as in the previous section, we get $\eta_{0} = \vec{b'}\cdot\vec{s'}/c$. Given a certain $u\varv$ length, the delay $\eta_0$ due to that source $\vec{s'}$ has the maximum value for a $u\varv$ cell such that $\vec{b'}$ points in the direction of $\vec{s'}$. Thus, the maximum delay due to this source is given by $\eta_0 = |\vec{b'}||\vec{s'}|/c$ and the wedge equation under a flat-sky approximation is given by: 
\begin{equation}\label{eq:wedge_flat_psi}
k_{\parallel} = k_{\perp}\dfrac{D_M(z)\mathcal{H}_0 E(z)}{c(1+z)}|\vec{s'}| = k_{\perp}\dfrac{D_M(z)\mathcal{H}_0 E(z)}{c(1+z)} \sin\psi.
\end{equation}
Setting $\psi=\pi/2$ for the horizon (maximum $k_{\parallel}$), it follows that:
\begin{equation}\label{eq:wedge_flat}
k^{\mathrm{flat}}_{\parallel} = k_{\perp}\dfrac{D_M(z)\mathcal{H}_0 E(z)}{c(1+z)}.
\end{equation}
This equation has conventionally been used to define the horizon lines in cylindrical power spectra of 21-cm cosmology. However, setting $\psi=\pi/2$ in Eq.~(\ref{eq:wedge_flat_psi}) violates the flat-sky approximation that was used to derive the equation in the first place. As a result, the equation incorrectly suggests that the maximum $k_{\parallel}$ value occurs at $\psi = \pi/2$, which does not correspond to the physical horizon, except when the phase center points exactly toward the zenith. For zenith phasing (a snapshot drift scan observation) with an array lying on the Earth's tangent plane, all $W$ coordinates are zero. So even without imposing the flat-sky approximation, Eq.~(\ref{eq:vcz_nopq}) reduces to Eq.~(\ref{eq:vcz_flat}). We note that for the special case of a zenith phasing, $a_0 = \pi/2$, and the LST-dependent horizon line derived in this analysis (Eq.~(\ref{eq:horizon_line_lst})) reduces to the conventional horizon line equation (Eq.~(\ref{eq:wedge_flat})).

\begin{figure*}
    \includegraphics[width=\hsize]{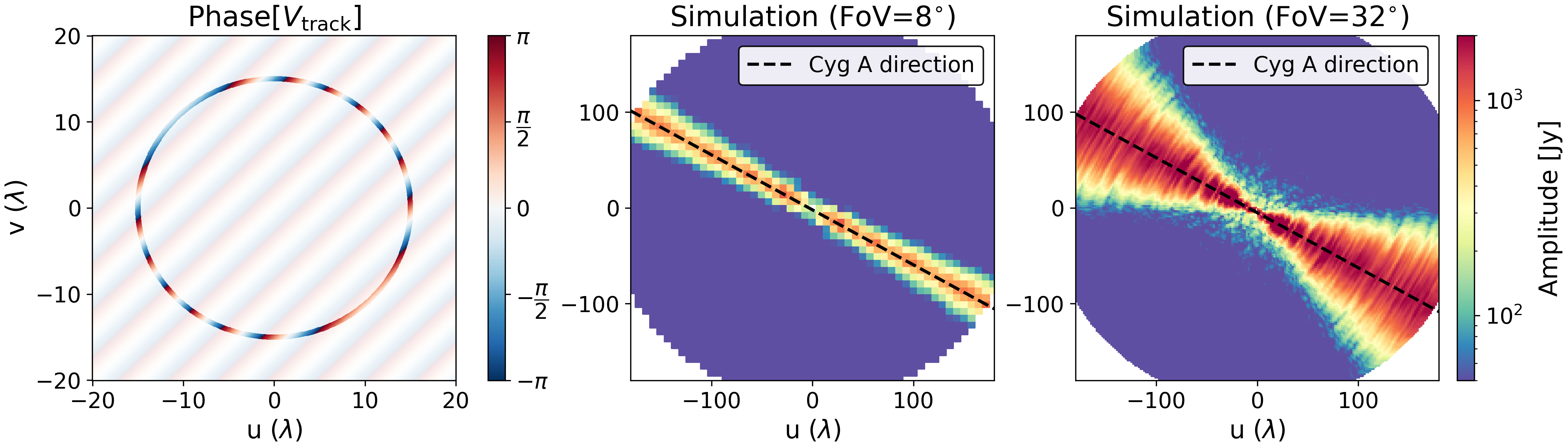}
    \caption{The signature of a source on the $u\varv$ plane. The left panel is a schematic diagram illustrating the quasi-sinusoidal behavior of the phase of visibilities due to a source along a circular $u\varv$ track. The middle and right panels show the visibility amplitude in the $u\varv$ plane for simulations of Cyg A with the phase center at the NCP, with FoVs of $8^{\circ}$ and $32^{\circ}$ respectively.}
    \label{fig:source_signature_2d}
\end{figure*}
\section{Source wedge in the cylindrical power spectrum}\label{sec:signature_source}
The horizon line is useful to understand the maximum extent of foreground contamination in the data. However, in many situations, the sky might be dominated by a few very strong sources far from the phase center. It is thus very useful to be able to predict the impact of a specific source on the cylindrical power spectrum. In this section, we derive an expression for the Fourier modes most strongly affected by a given source in the cylindrical power spectrum.

\subsection{Effect of sampling and gridding}\label{sec:sampling_gridding}
Under the flat-sky approximation, the delay due to a source lying in the plane perpendicular to the LoS ($\vec{s'}$) is given by $\eta_0 = \vec{b'}\cdot\vec{s'}/c$ (Sect.~\ref{sec:drift_scan}). For the horizon delay, we used the maximum value of $\eta_0$, which occurs when $\vec{b'}$ and $\vec{s'}$ point in the same direction. However, in this section, the aim is not to determine the maximum delay caused by a source, but rather to identify the delay at which the source has the maximum power. This is important for analyzing the cause of artifacts in the cylindrical power spectrum. If $\beta$ is the angle between $\vec{b'}$ and $\vec{s'}$, all delays between zero (for $\beta=\pi/2$) and $|\vec{b'}||\vec{s'}|/c$ (for $\beta=0$) have the same power. Thus, the signature of a single source at an angle $\psi$ from the phase center is a wedge with equal power along lines of $k_{\parallel} \propto k_{\perp}$, with the maximum slope given by Eq.~(\ref{eq:wedge_flat_psi}).

However, this simple situation is only applicable to an interferometer with perfect $u\varv$ coverage. An imperfect $u\varv$ coverage gives rise to a point spread function (PSF) that is not a delta function. If the $u\varv$ tracks are circular, the PSF sidelobes also have a circular symmetry. Given a certain FoV within which we make the image, the PSF sidelobes of sources far from the phase center run through the field. For sources sufficiently far away compared to the FoV, the signature of the sidelobes in the $u\varv$ plane approaches a straight line of higher source power along the direction of the source, that is, perpendicular to the sidelobe ripples. Thus, instead of uniform power in the $u\varv$ plane, we end up with higher power in the $u\varv$ plane in the direction of the source (as illustrated in the middle and right panels of Fig.~\ref{fig:source_signature_2d}).

In the $u\varv$ domain, we can explain this as follows. The visibility due to a source is a complex exponential in the $u\varv$ plane and an interferometer samples this visibility along $u\varv$ tracks. These sampled visibilities are gridded on the $u\varv$ plane with a convolution kernel whose width is decided by the target FoV. Now the visibility (both real and imaginary parts) along a given $u\varv$ track has slower fluctuations in the direction of the source and faster fluctuations in the perpendicular direction. The left panel of Fig.~\ref{fig:source_signature_2d} is a schematic diagram of the phase of the visibilities due to a source at a certain angular distance from the phase center located at the bottom-right corner of the image, after sampling at a circular $u\varv$ track. We highlight this track and the $u\varv$ cells not sampled are shown with faded colors. Here we can see a quasi-sinusoidal behavior of the sampled visibility as a function of the angle along the $u\varv$ track, with the slowest variations in the direction of the source. Thus, a convolution kernel of a given width suppresses the amplitude of the sampled visibility more in the direction perpendicular to the source direction \citep{offringa2012post}. 

The middle and right panels of Fig.~\ref{fig:source_signature_2d} show the visibility amplitude in the $u\varv$ plane for a $12\,$h simulated observation of Cygnus A (Cyg A) with the phase center at the NCP. We see that the $u\varv$ plane has higher power in the direction of Cyg A (lying toward the bottom right of the image). The gridded $u\varv$ plane is constructed using only the data up to a maximum FoV from the phase center. The visibility amplitudes in the $u\varv$ plane for FoVs of $8^{\circ}$ (middle panel) and $32^{\circ}$ (right panel) show that the signature of the source in the $u\varv$ plane is wider for a larger FoV. This is because a smaller FoV leads to a wider convolution kernel, that suppresses fluctuations as we move slightly away from the source direction. For a large number of $u\varv$ tracks, this results in a sharper feature in the $u\varv$ plane compared to that of a wider FoV. This effect is also discussed in Appendix~\ref{sec:sampling_gridding_analytical}, where we illustrate the effect of sampling and gridding analytically, for a circular $u\varv$ track. It should be noted that the line of highest power is oriented exactly along the source direction only for circular $u\varv$ tracks because the real and imaginary parts of the visibility along a $u\varv$ track have the slowest fluctuations exactly in the source direction only if the $u\varv$ track is circular. For elliptical $u\varv$ tracks, for example, the slowest fluctuations will not occur exactly along the source direction. This results in a line of highest power that is not aligned along the source direction, with the extent of the deviation being higher for ellipses with higher eccentricity. Additionally, the functional form of the gridding kernel affects the direction of maximum power in the $u\varv$ plane. Both these effects are discussed in Appendix~\ref{sec:sampling_gridding_analytical}.

\subsection{Delay due to a source}
In the previous section, we discussed how the sampling and gridding performed in the process of measurement by an interferometer and construction of gridded visibility cubes results in a higher amplitude of the gridded visibilities along the source direction in the $u\varv$ plane. We now derive an expression for the delay at which this higher amplitude occurs. Working in the same XYZ coordinate system as Sect.~\ref{sec:analytical}, the delay due to a source located at $H,\delta$ is given by Eq.~(\ref{eq:delay_arbitrary}). In the absence of the exact $u\varv$ coverage, throughout this section and in the entire derivation of source line equations, we make the simplifying assumption that the $u\varv$ tracks are circular. Thus, the highest power is due to a baseline pointing in the direction of the source from the phase center (i.e., along $\hat{\vec{s}}-\hat{\vec{p}}$)\footnote{Incidentally, this is also the baseline orientation that results in the maximum delay due to a given source. This is verified by numerical maximization and described in Appendix~\ref{sec:source_delay_validation}.}. Since we assume that the baselines lie on the tangent plane of the Earth, the expression for $\vec{b}\cdot(\hat{\vec{s}}-\hat{\vec{p}})$ for such a baseline is given by the baseline length multiplied by the length of the projection of $(\hat{\vec{s}}-\hat{\vec{p}})$ on the plane perpendicular to $\hat{\vec{z}}$. The corresponding delay ($\eta_0=\vec{b}\cdot(\hat{\vec{s}}-\hat{\vec{p}})/c$) is given by:
\begin{align}\label{eq:source_delay}
\eta_0 =& \dfrac{|\vec{b}|}{c}\,|(\hat{\vec{s}}-\hat{\vec{p}}) - ((\hat{\vec{s}}-\hat{\vec{p}})\cdot\hat{\vec{z}})\hat{\vec{z}}|\nonumber\\
=& \dfrac{|\vec{b}|}{c}\,\biggl((\cos \delta_0 \sin H_0-\cos \delta \sin H)^2+\bigl(\sin \phi(\cos \delta_0 \cos H_0\nonumber\\&
-\cos \delta \cos H)+\cos \phi(\sin \delta-\sin \delta_0)\bigr)^2\biggr)^{1/2}.
\end{align}

\subsection{Source line in the cylindrical power spectrum}\label{sec:source_ps}
We next derive the equation describing the line in the cylindrical power spectrum due to a given source. From Eq. (\ref{eq:delay_source_noflat}), the delay in a $u\varv$ cell without imposing the flat-sky approximation is $\vec{b}\cdot(\hat{\vec{s}}-\hat{\vec{p}})/c$. Inserting the expression for the delay at which a specific source $\hat{\vec{s}}$ contributes the maximum power (from Eq.~(\ref{eq:source_delay})), we get:
\begin{align}\label{eq:source_delay_1}
    \eta^{\mathrm{source}}_{0} = \dfrac{|\vec{b}|}{c}\,|(\hat{\vec{s}}-\hat{\vec{p}}) - ((\hat{\vec{s}}-\hat{\vec{p}})\cdot\hat{\vec{z}})\hat{\vec{z}}|
    \equiv \dfrac{|\vec{b}||\vec{a}|}{c},
\end{align}
where $\vec{a} = (\hat{\vec{s}}-\hat{\vec{p}}) - ((\hat{\vec{s}}-\hat{\vec{p}})\cdot\hat{\vec{z}})\hat{\vec{z}}$. It should be noted that $\vec{a}$ points in the direction of the baseline responsible for the maximum power. Now, the projected length of the baseline $\vec{b}$ on the plane perpendicular to the phase center is given by $|\vec{b'}|$ = $|\vec{b}|\sin\gamma$, where $\gamma$ is the angle between $\vec{b}$ and $\hat{\vec{p}}$. Since $\vec{b}$ points along $\hat{\vec{a}} = \vec{a}/|\vec{a}|$, the relation between $|\vec{b'}|$ and $|\vec{b}|$ becomes:
\begin{equation}\label{eq:source_delay_2}
    |\vec{b'}| = |\vec{b}|\sqrt{1-\biggl(\dfrac{\vec{a}}{|\vec{a}|}\cdot\hat{\vec{p}}\biggr)^2}.
\end{equation}
Combining Eqs. (\ref{eq:source_delay_1}) and (\ref{eq:source_delay_2}), we get:
\begin{align}\label{eq:source_special_case}
\eta^{\mathrm{source}}_{0} = \dfrac{\sqrt{u^2+\varv^2}}{\nu_0}\,\left(\dfrac{|\vec{a}|^2}{\sqrt{|\vec{a}|^2-(\vec{a}\cdot\hat{\vec{p}})^2}}\right).
\end{align}
Finally, converting to cosmological coordinates using Eq.~(\ref{eq:conversion}) and expanding the expressions for the vectors in terms of $H,\delta,H_0,\delta_0$, and $\phi$, the source line equation in the cylindrical power spectrum takes the form:
\begin{align}\label{eq:source_line}
k^{\mathrm{source}}_{\parallel} &= k_{\perp}\dfrac{D_M(z)\mathcal{H}_0 E(z)}{c(1+z)}\nonumber\\
\times&{\tiny\dfrac{\begin{aligned} &  (\sin{H} \cos{\delta} - \sin{H_0} \cos{\delta_0} )^{2} +  (\sin {\phi} \cos{\delta} \cos{H} \\ & - \sin {\phi} \cos{\delta_0} \cos{H_0} - \sin{\delta} \cos {\phi} + \sin{\delta_0} \cos {\phi} )^{2}\end{aligned}}{\sqrt{
  \begin{aligned}
&  (\sin{H} \cos{\delta} - \sin{H_0} \cos{\delta_0} )^{2} -  ( (\sin{H} \cos{\delta} - \sin{H_0} \cos{\delta_0} ) \\ & \sin{H_0} \cos{\delta_0} +  (\sin {\phi} \cos{\delta_0} \cos{H_0} - \sin{\delta_0} \cos {\phi} )  (\sin {\phi} \cos{\delta} \cos{H} \\ & - \sin {\phi} \cos{\delta_0} \cos{H_0} - \sin{\delta} \cos {\phi} + \sin{\delta_0} \cos {\phi} ) )^{2} +  (\sin {\phi} \cos{\delta} \\ & \cos{H} - \sin {\phi} \cos{\delta_0} \cos{H_0} - \sin{\delta} \cos {\phi} + \sin{\delta_0} \cos {\phi} )^{2}
\end{aligned}
}}}.
\end{align}
By substituting $H=t-\alpha$ and $H_0=t-\alpha_0$, we get the expression of the line due to a given source in the cylindrical power spectrum as a function of LST, for a particular phase center. In Sect.~\ref{sec:sim_source}, we verify the validity of this equation against representative simulations for a range of observation parameters. We note that in Eq.~(\ref{eq:source_line}), the slope of the line in the ($k_{\perp},k_{\parallel}$) space is a function of LST. Hence, given an observation duration, the line due to a source will assume different slopes at different times. This will lead to a wedge-like region in the cylindrical power spectrum most affected by a source during the observation. This will be illustrated in Sect. \ref{sec:sim_source}.

\subsection{Special case of zenith phasing}\label{sec:drift_scan_source}
Under a flat-sky approximation, the expression of the delay in a $u\varv$ cell due to a source $\vec{s'}$ is $\vec{b'}\cdot\vec{s'}/c$ (Sect. \ref{sec:drift_scan}). We showed in Sect.~\ref{sec:sampling_gridding} that the maximum power due to a source occurs on $u\varv$ cells that lie along the direction of $\vec{s'}$. Thus $\eta_0^{\mathrm{source}} = |\vec{b'}||\vec{s'}|/c$ and the signature of the highest power due to a source under the flat-sky approximation is given by:
\begin{align}\label{eq:source_line_flat}
k^{\mathrm{source}}_{\parallel} = k_{\perp}\dfrac{D_M(z)\mathcal{H}_0 E(z)}{c(1+z)} \sin\psi.
\end{align}
However, as in the case of the horizon line equation with flat-sky approximation (Eq.~(\ref{eq:wedge_flat})), this equation breaks down for sources far away from the phase center where the condition $\psi \rightarrow 0$ does not hold.

As discussed in Sect. \ref{sec:drift_scan}, the special case of zenith phasing is equivalent to imposing a flat-sky approximation. When the phase center is at the zenith, $\hat{\vec{z}}=\hat{\vec{p}}$. If $\psi$ is the angle between $\hat{\vec{s}}$ and $\hat{\vec{p}}$, $\vec{a} = \hat{\vec{s}} - \hat{\vec{p}}\cos\psi$. It follows that $|\vec{a}| = \sin\psi$ and $\vec{a}\cdot\hat{\vec{p}} = 0$.
Inserting these expressions in the source line equation derived in this analysis (Eq.~(\ref{eq:source_special_case})) and converting to cosmological coordinates, we get back the conventional source line equation (Eq.~(\ref{eq:source_line_flat})).

\begin{figure}
    \includegraphics[width=\hsize]{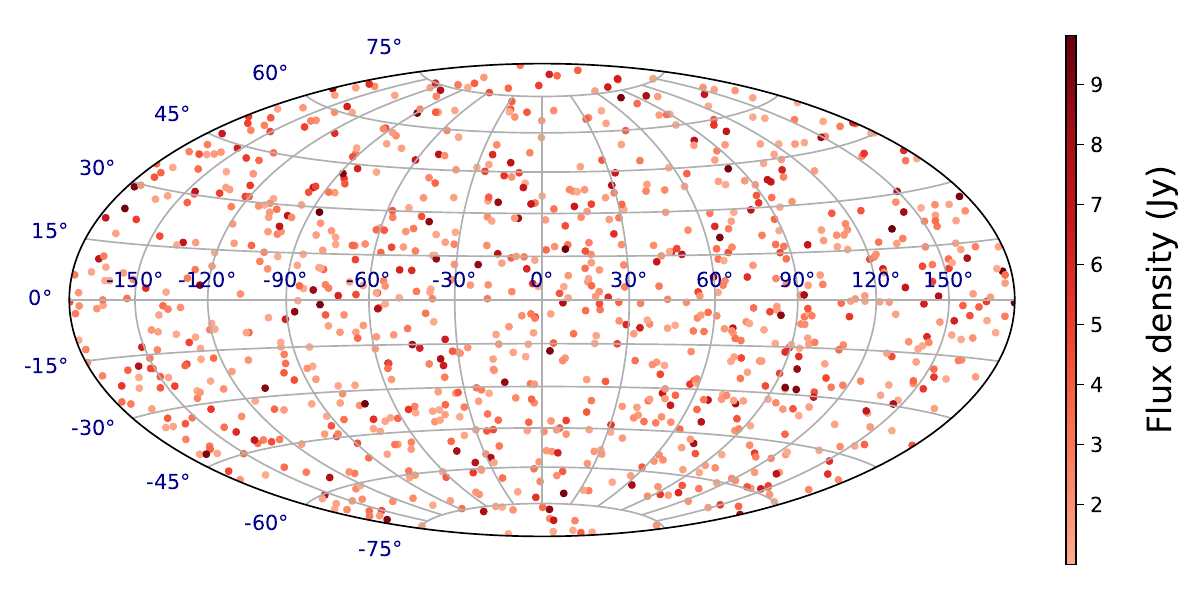}\\
    \includegraphics[width=\hsize]{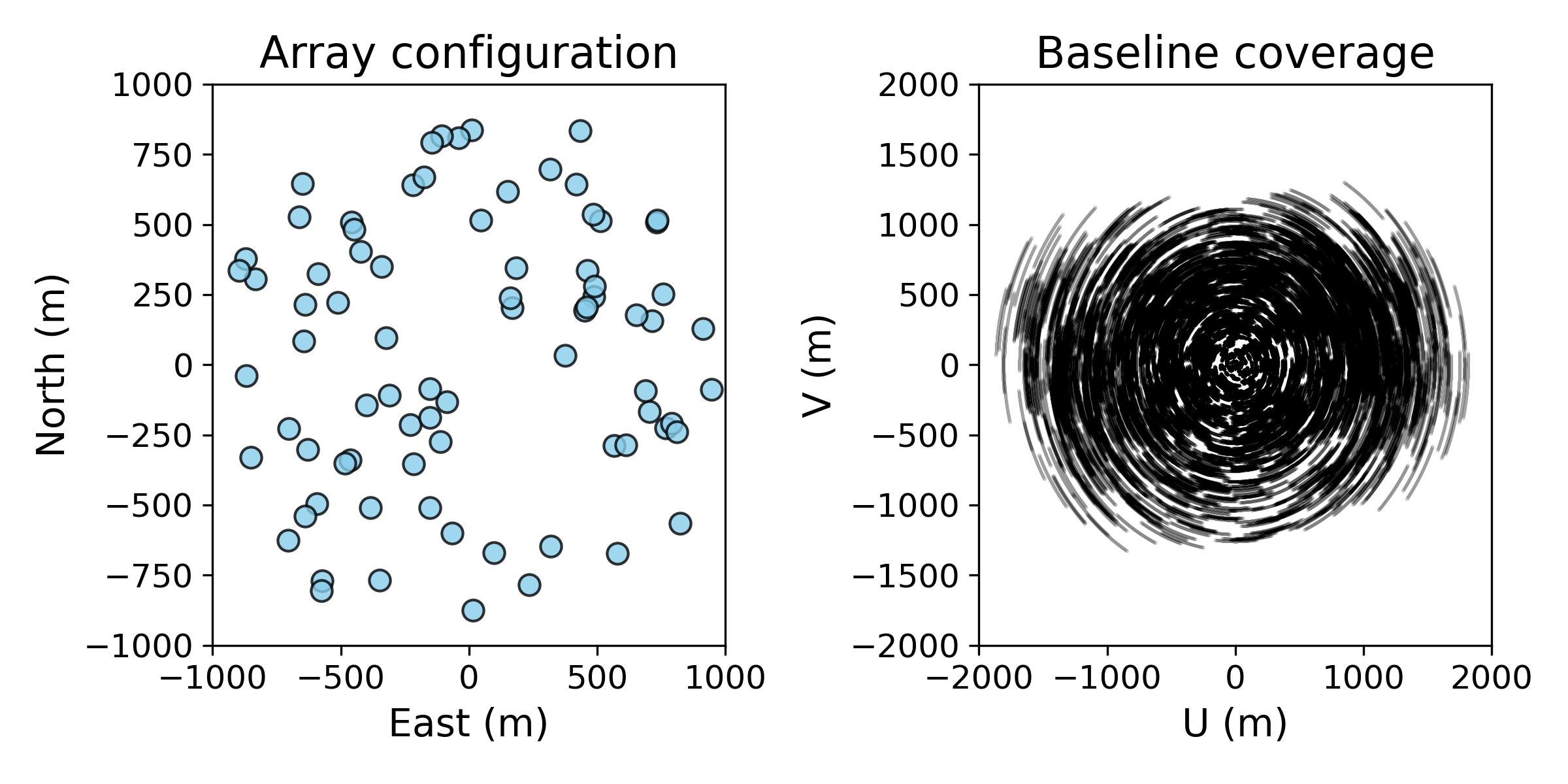}
    \caption{The inputs to the simulations used for validating the derived equations. Top: The sky model used in the simulations performed for validating the horizon line equations. Bottom left: The local array configuration. Bottom right: An example baseline coverage, for a $1\,$h simulation with NCP phasing with the array placed at the location of NenuFAR.}
    \label{fig:sim_inputs}
\end{figure}
\section{Simulations}\label{sec:simulations}
In Sect.~\ref{sec:horizon_ps}, we derived two equations (Eqs. (\ref{eq:horizon_line_lst}) and (\ref{eq:horizon_line_full})) which describe the horizon line in the cylindrical power spectrum at a given LST and for full synthesis. In Sect.~\ref{sec:signature_source}, we derived an equation that describes the source line in the cylindrical power spectrum at a given LST (Eq.~(\ref{eq:source_line})). These equations are dependent on several parameters, such as the latitude of the telescope, the coordinates of the phase center, and the time of observation. We now verify the validity of these equations over these parameters through a set of representative simulations for a range of observation parameters and a set of latitudes and longitudes of current and upcoming radio telescopes.

\subsection{Horizon line}\label{sec:sim_horizon}
In this section, we verify both Eqs. (\ref{eq:horizon_line_lst}) and (\ref{eq:horizon_line_full}) against full-sky simulations. The sky model contains 1000 sources uniformly distributed in the sky with the flux ($S$) of the sources described by a flux density distribution $dN/dS \propto S^{-1.54}$ \citep{franzen2016154} between 1 and 10 Jy, with each component having a $\nu^{-0.8}$ frequency dependence. This model is sufficient to assess the effect of sources from all over the sky down to the horizon for any location on the Earth, at any LST. The sky model used in the simulations is shown in the top panel of Fig.~\ref{fig:sim_inputs}. We constructed a fiducial array that (locally) has a random distribution of 79 stations in a 2 km diameter region, with the vertical positions having a jitter of $\pm 2$ m. This array was then placed at the geographical locations of NenuFAR, LOFAR, and SKA. The array configuration and the baseline coverage (for NCP phasing with the array placed at the location of NenuFAR) are shown in the bottom panels of Fig.~\ref{fig:sim_inputs}.

\begin{figure*}
    \includegraphics[width=\hsize]{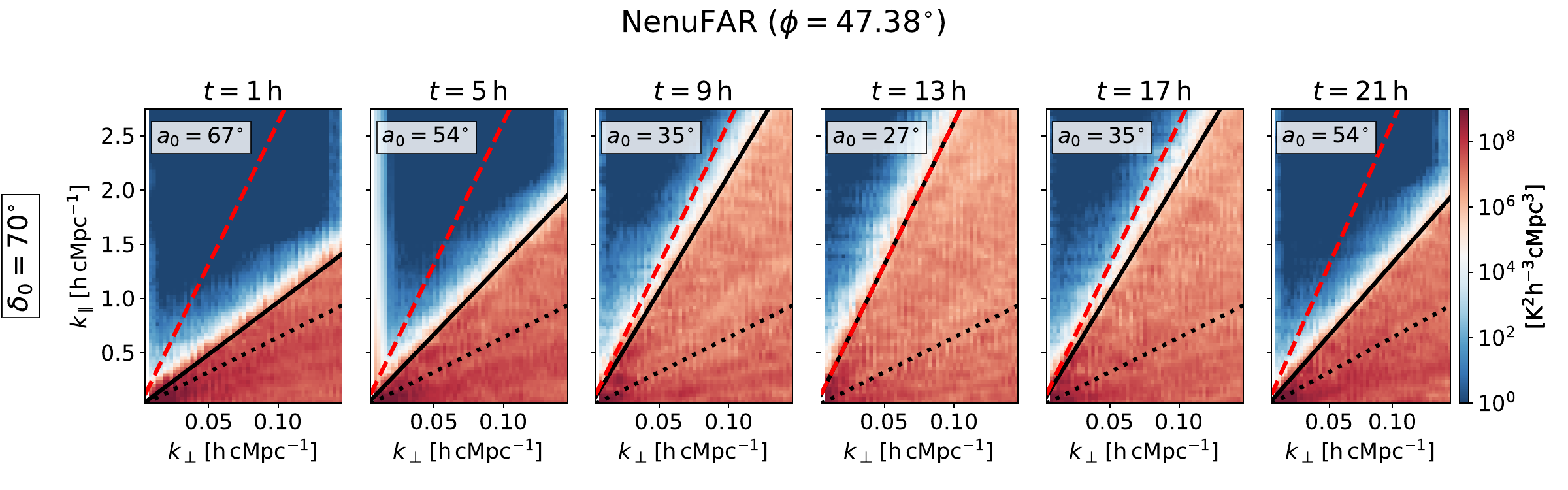}
    \includegraphics[width=\hsize]{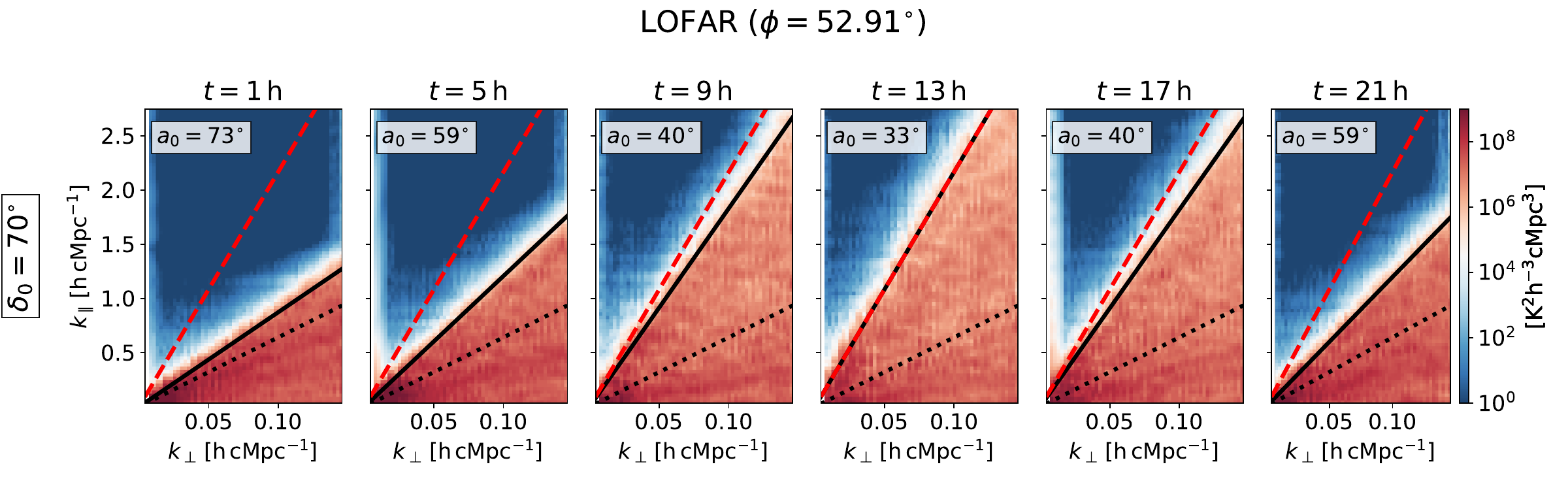}
    \includegraphics[width=\hsize]{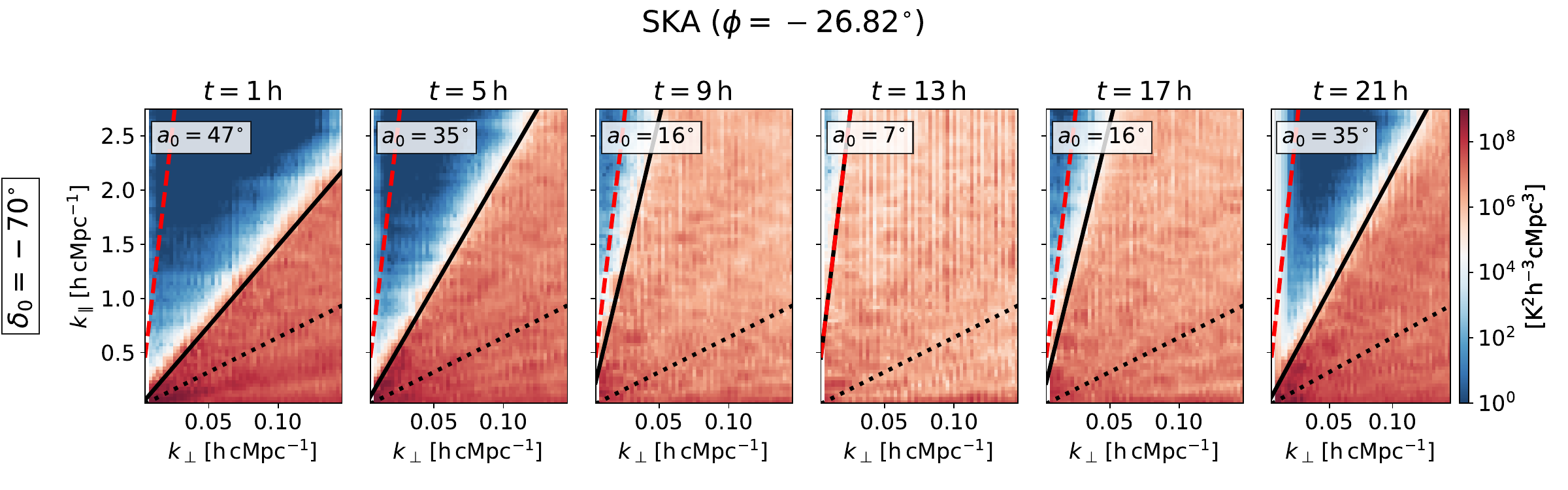}
    \caption{Cylindrical power spectra for full-sky simulations with the array placed at the locations of NenuFAR, LOFAR, and SKA corresponding to a phase center declination of $70^{\circ}$ ($-70^{\circ}$ for SKA). The different columns correspond to different LSTs at which the $1\,$h simulations are centered. The dotted line is the conventional horizon line. The black solid line is the LST-dependent horizon line (Eq.~(\ref{eq:horizon_line_lst})). The red dashed line is the horizon line for full synthesis (Eq.~(\ref{eq:horizon_line_full})). $a_0$ is the altitude of the phase center at the LST at which the simulation is centered. The results of the remaining simulations for the celestial pole and zenith transit phase centers are shown in Fig. \ref{fig:horizon_appx}.}
    \label{fig:horizon}
\end{figure*}

The simulations were performed using \texttt{OSKAR} \citep{dulwich_2020_3758491}. Since the aim is to test the derived equations that describe the maximum extent of foregrounds without considering primary beam effects, we disabled station beam effects and used \texttt{``station type = Isotropic beam''} for all simulations. In these simulations, the visibility of a source is not predicted when it goes below the physical horizon, as would be expected in an actual observation. The frequency range was chosen to be 61 to 72 MHz with 177 channels and the time resolution was chosen to be 4 s\footnote{These parameters are the same as those used by \cite{munshi2024first}. However, the exact values of the frequency range and time resolution are not relevant for the purpose of these simulations.}. We performed $1\,$h simulations centered at six LSTs each separated by $4\,$h (and hence spanning a day). We repeated the simulations for three phase centers and chose the phase center RA to be $1\,$h such that the phase center crosses the local meridian at t=$1\,$h and thus at the maximum altitude. It should be noted that since a visibility and its complex conjugate are sampled at $u,\varv$ and $-u,-\varv$ respectively, 12 hours is sufficient to yield a complete $u\varv$ coverage. However, the situations separated by $12\,$h are not identical since the phase center can be at different elevations, thus requiring different geometric phasing. The settings used for the simulations for validating the horizon lines are summarized in the top section of Table~\ref{tab:sim_params}.

The power spectra were constructed using the power spectrum pipeline \texttt{pspipe}\footnote{\url{https://gitlab.com/flomertens/pspipe}}, which is used for estimating the power spectra for the NenuFAR and LOFAR 21-cm cosmology projects. \texttt{pspipe} uses \texttt{WSClean} \citep{offringa2014wsclean} to construct gridded dirty image cubes. A Hann filter with a FoV of 8 degrees was applied in the spatial direction and a Blackman-Harris filter along the frequency direction of the dirty image cube before Fourier transformation into the $u\varv\eta$ space. Finally, the modulus squared of the gridded data cube in the Fourier space was used to construct the cylindrical power spectrum.

Figure~\ref{fig:horizon} shows the cylindrical power spectra for the simulations corresponding to NenuFAR, LOFAR, and SKA for a phase center declination of $70^{\circ}$ ($-70^{\circ}$ for SKA). The results of the remaining simulations corresponding to the celestial pole and zenith transit phase centers are shown in Fig. \ref{fig:horizon_appx}. For each situation, both the LST-dependent line (Eq.~(\ref{eq:horizon_line_lst}): solid black line) and the full synthesis line (Eq.~(\ref{eq:horizon_line_full}): red dashed line) are shown. The LST-dependent horizon line (Eq.~(\ref{eq:horizon_line_lst})) will in general move through the ($k_{\perp},k_{\parallel}$) space slightly during the $1\,$h observation. So the horizon $k_{\parallel}$ is evaluated at a series of 100 equally spaced time points within the 1 h observation duration and the line corresponding to the maximum $k_{\parallel}$ is shown. We see that both the LST-dependent line and the full synthesis line perfectly predict the full sky horizon limit in all situations in both Figs.~\ref{fig:horizon} and \ref{fig:horizon_appx}. For each row, the full synthesis line indicates the maximum possible slope of the LST-dependent horizon line, that is, the horizon line if we were to perform a full synthesis. The altitude of the phase center ($a_0$) at the LST at which the simulation is centered is mentioned on each plot. We see that the horizon line is steeper when the phase center goes closer to the horizon, as predicted by Eq.~(\ref{eq:horizon_line_lst}). The horizon extent is lower for the LOFAR simulations compared to the NenuFAR simulations. This is because LOFAR is at a higher latitude, and the same phase centers are closer to the zenith compared to NenuFAR. The conventional horizon line (Eq.~(\ref{eq:wedge_flat}): black dotted line) fails to correctly describe the full sky foreground wedge horizon except in situations where the phase center is very close to the zenith (bottom-left panel for each location in Fig. \ref{fig:horizon_appx}). In these situations, the LST-dependent horizon line equation (Eq.~(\ref{eq:horizon_line_lst})) reduces to the conventional horizon line equation (Eq.~(\ref{eq:wedge_flat})), as described in Sect.~\ref{sec:drift_scan}. Also, the conventional equations predict that the horizon line is independent of time, the coordinates of the phase center, and the latitude of the telescope. However, in Figs.~\ref{fig:horizon} and \ref{fig:horizon_appx}, we can see that this is not the case for full-sky simulations with an arbitrary phase center, where the extent of the foreground wedge is highly dependent on these parameters.

\begin{table}[]
\caption{The settings for which the simulations were repeated, to verify the horizon line equations and the source line equation over a range of observation parameters.}
\label{tab:sim_params}
\centering
\setlength{\tabcolsep}{2pt}
\resizebox{\hsize}{!}{
\begin{tabular}{@{}clccccclccccc@{}}
\toprule
\multicolumn{13}{c}{Horizon Line}                                                                                                                                          \\ \midrule
\multicolumn{1}{c|}{Location}       & \multicolumn{6}{l|}{NenuFAR, LOFAR}                                        & \multicolumn{6}{l}{SKA}                                 \\
\multicolumn{1}{c|}{$\delta_0$}   & \multicolumn{6}{l|}{NCP, $70^{\circ}$, Zenith transit}                              & \multicolumn{6}{l}{SCP, $-70^{\circ}$, Zenith transit}           \\
\multicolumn{1}{c|}{$\alpha_0$}    & \multicolumn{6}{l|}{$1\,$h}                                                    & \multicolumn{6}{l}{$1\,$h}                                  \\
\multicolumn{1}{c|}{$t_{\text{center}}$ (h)} & \multicolumn{6}{l|}{1, 5, 9, 13, 17, 21}                                   & \multicolumn{6}{l}{1, 5, 9, 13, 17, 21}                 \\
\multicolumn{1}{c|}{Span (h)}   & \multicolumn{6}{l|}{1}                                                     & \multicolumn{6}{l}{1}                                   \\ \midrule
\multicolumn{13}{c}{Source Line}                                                                                                                                           \\ \midrule
\multicolumn{1}{c|}{Location}       & \multicolumn{6}{l|}{NenuFAR, LOFAR}                                        & \multicolumn{6}{l}{SKA}                                 \\
\multicolumn{1}{c|}{Source}         & \multicolumn{6}{l|}{Cyg A, Cas A}                                          & \multicolumn{6}{l}{Vir A, Cen A}                        \\
\multicolumn{1}{c|}{$\delta_0$}   & \multicolumn{6}{l|}{NCP, Zenith transit}                                  & \multicolumn{6}{l}{SCP, Zenith transit}                         \\
\multicolumn{1}{c|}{$\alpha_0$}    & \multicolumn{6}{l|}{$1\,$h}                                                    & \multicolumn{6}{l}{13h}                                 \\
\multicolumn{1}{c|}{$t_{\text{start}}$ (h)}  & \multicolumn{1}{c}{0.5} & 4.5 & 8.5 & 0.5 & 0.5 & \multicolumn{1}{c|}{0.5} & \multicolumn{1}{c}{8.5} & 12.5 & 16.5 & 8.5 & 8.5 & 8.5 \\
\multicolumn{1}{c|}{Span (h)}       & \multicolumn{1}{c}{1}   & 1   & 1   & 4   & 8   & \multicolumn{1}{c|}{12}  & \multicolumn{1}{c}{1}   & 1    & 1    & 4   & 8   & 12  \\ \bottomrule
\end{tabular}
}
\end{table}

\begin{figure*}
    \includegraphics[width=\hsize]{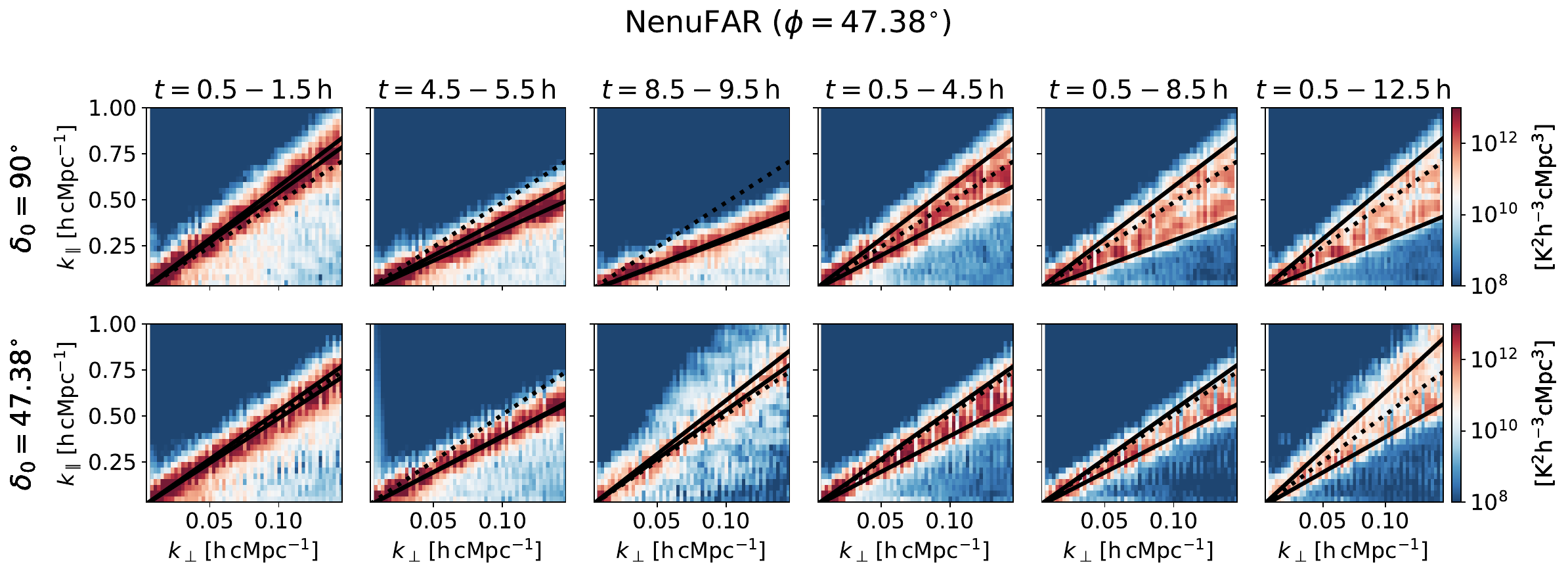}
    \includegraphics[width=\hsize]{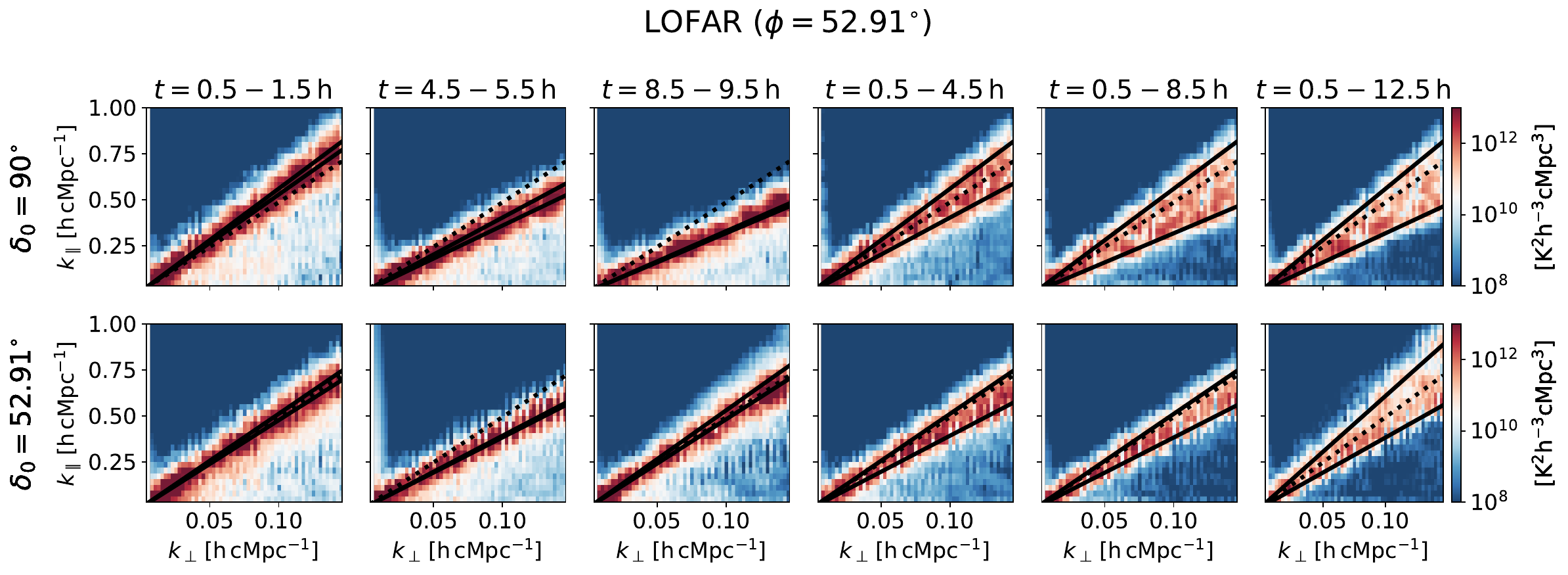}
    \includegraphics[width=\hsize]{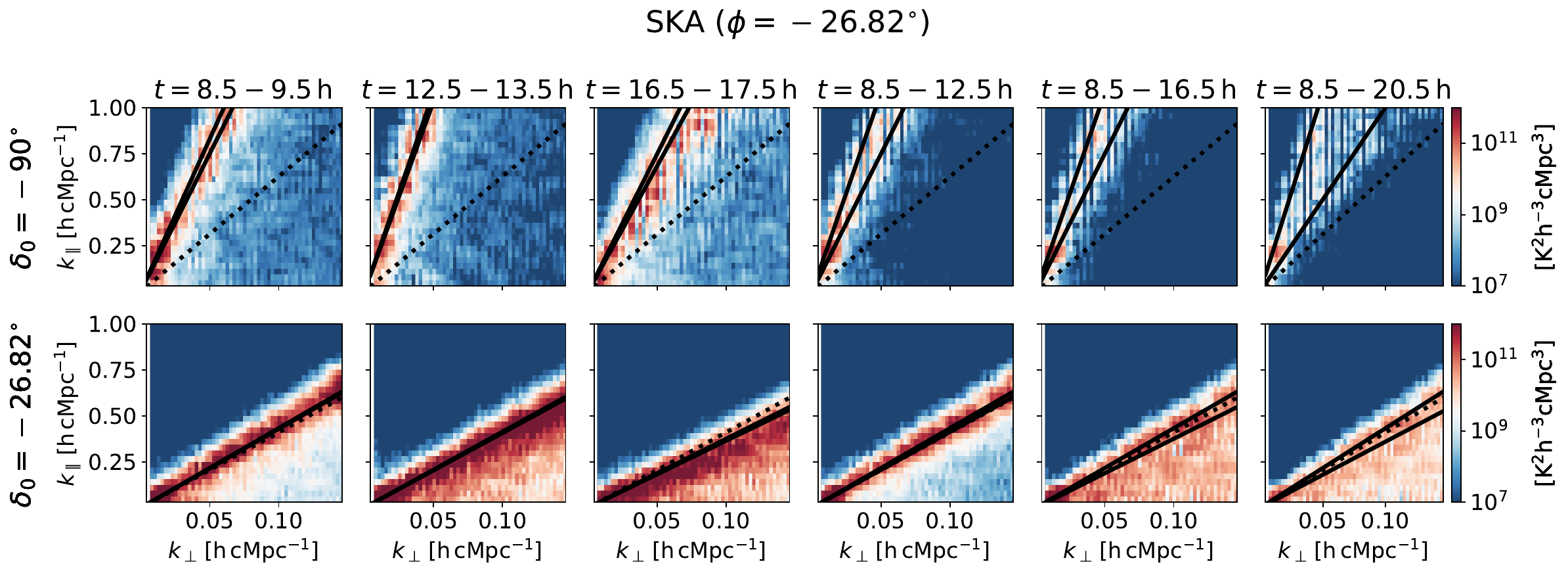}
    \caption{Cylindrical power spectra for Cyg A simulations at the locations of NenuFAR and LOFAR, and Vir A simulations at the location of SKA. For each location, the two rows correspond to two phase center declinations ($\delta_0$) of the celestial pole (upper row) and zenith transit (lower row). The different columns correspond to the different LST ranges for which the simulations were repeated. The dotted line is the conventional source line. The black solid lines correspond to the maximum and minimum source lines for the observation duration, obtained from the source line equation derived in this analysis (Eq.~(\ref{eq:source_line})). The remaining Cas A and Cen A simulation results are shown in Fig. \ref{fig:source_appx}.}
    \label{fig:source}
\end{figure*}

\subsection{Source line}\label{sec:sim_source}
In this section, we verify Eq.~(\ref{eq:source_line}) against simulations of a single source for a range of observation parameters. In the simulations, we used the same random array at the locations of NenuFAR, LOFAR, and SKA. The simulation and power spectrum generation were performed in the same manner as described in Sect.~\ref{sec:sim_horizon}, except here the sky model consists of a single delta function at the position of a chosen source. Also, here we performed simulations for six LST ranges: three of them being $1\,$h observations at different LSTs separated by $4\,$h, and the other three being observations of different durations with the same start LST. This allows us to study how a longer observation duration results in a range of modes in the power spectrum getting affected by a strong source. The simulations were repeated for two sources, two phase centers, and three array locations corresponding to NenuFAR, LOFAR, and SKA. We chose Cyg A and Cassiopeia A (Cas A) for NenuFAR and LOFAR because they have the strongest impact on the data of these telescopes. For SKA, we chose Virgo A (Vir A) and Centaurus A (Cen A) and a phase center RA of 13 h to ensure that they remain above the horizon for all the observation durations considered. The settings for which the simulations for validating the source line were repeated are summarized in the bottom section of Table~\ref{tab:sim_params}.

Figure~\ref{fig:source} shows the cylindrical power spectra for the simulations used for validating the source line equations corresponding to Cyg A (for NenuFAR and LOFAR) and Vir A (for SKA). The remaining simulations of Cas A (for NenuFAR and LOFAR) and Cen A (for SKA) are shown in Fig. \ref{fig:source_appx}. In each situation, the source line is evaluated from Eq.~(\ref{eq:source_line}) at 100 equally spaced time samples and the maximum and minimum source lines for the observation duration are shown (in solid black lines). The dotted black lines indicate the conventional source line derived with a flat-sky approximation (Eq.~(\ref{eq:source_line_flat})). We see that in all situations, the new source lines provide a better depiction of the range of modes most strongly affected by a source in the cylindrical power spectrum. The conventional source lines only depend on the angular separation between the source and the phase center, and hence do not move with LST (i.e., along a row in Figs.~\ref{fig:source} and \ref{fig:source_appx}). On the other hand, the new equations capture the trend of the movement of the source lines in the ($k_{\perp},k_{\parallel}$) space. For longer observations of 4, 8, and 12 h (the three right-most columns in Figs.~\ref{fig:source} and \ref{fig:source_appx}) the new equations can describe the range of modes affected by a source during a synthesis observation.

An important distinction between the horizon line and the source line equations is that the horizon line equations (both Eqs. (\ref{eq:horizon_line_lst}) and (\ref{eq:horizon_line_full})) have the maximum delay in the numerator (from Eqs. (\ref{eq:horizon_arbitrary_lst}) and (\ref{eq:horizon_arbitrary})) divided by the projection of the baseline responsible for the maximum delay on the $u\varv$ plane, in the denominator. Since the baseline responsible for the maximum delay lies on the plane containing the zenith and the phase center, this is also the baseline which has the smallest projection on the $u\varv$ plane. This means that the numerator is the maximum possible delay and the denominator is the minimum possible projection, thus resulting in the maximum possible extent of power in the cylindrical power spectrum. However, for the source line equation (Eq.~(\ref{eq:source_line})), though the numerator corresponds to the maximum delay due to the source (see Appendix~\ref{sec:source_delay_validation}), the denominator does not necessarily correspond to the minimum possible projection of a baseline on the $u\varv$ plane. Thus, it is not surprising to have modes above the maximum source line that are affected by the power due to the source in some of the cases. Also, note that the power does not go to zero below the minimum source lines, since baselines oriented in other directions also contribute power, even though it is attenuated due to sampling and gridding (as described in Sect.~\ref{sec:sampling_gridding}).

In some situations, we see that the peak power extends beyond the source lines. This is possible because, in the derivation of the source line equation, we take the assumption that the maximum power in the $u\varv$ plane is in the direction of the source. However, the actual signature of a source in the $u\varv$ plane does not always have the highest power exactly in the source direction, and the exact functional form of the amplitude of the sampled gridded visibilities will depend on the shape of the $u\varv$ tracks and also on the nature of the convolution function (see Sect.~\ref{sec:sampling_gridding} and Appendix~\ref{sec:sampling_gridding_analytical}). Therefore the power due to the source slightly away from the source direction can end up slightly away from the predicted source line direction in the cylindrical power spectrum. This effect will be more pronounced for phase centers very close to the physical horizon. This is when the denominator of Eq.~(\ref{eq:source_line}) can have values very close to zero, and slight changes in the numerator will result in large variations of the $k_{\parallel}$ predicted by the source line equation. In Figs.~\ref{fig:source} and \ref{fig:source_appx}, we indeed see that the peak of the source line deviates from the predicted lines only for SKA simulations with the phase center at the SCP, which is very close to the physical horizon. It should be noted that phase centers that reach very close to the horizon are rarely chosen for 21-cm cosmology analyses since the sensitivity to the 21-cm signal decreases significantly due to attenuation by the dipole beam of receiving antennas. In Appendix~\ref{sec:source_line_specific}, we describe simulations that were performed for specific situations where the $u\varv$ tracks run along the source direction and illustrate its impact on the $u\varv$ plane and the power spectrum.

\section{Discussion}\label{sec:discussion}
The revised and new horizon and source line equations that we have developed accurately describe the full sky signature of foregrounds in the cylindrical power spectrum. In this section, we discuss the main limitations of the equations, their key characteristics, and their potential practical applications in 21-cm cosmology analyses. 
\subsection{Limitations}
The following are the main limitations of the horizon and source line equations derived in this analysis.
\paragraph{Impact of the primary beam on horizon lines:}The horizon line equations (Eqs. (\ref{eq:horizon_line_full}) and (\ref{eq:horizon_line_lst})) describe the maximum possible extent of foregrounds in the cylindrical power spectrum. However, we have not considered the impact of the primary beam of the interferometric elements in the derivation, which has a damping effect on the power far from the phase center and will decrease the extent of contamination at high $k_{\parallel}$, particularly so for instruments with a narrow primary beam extent. So while these lines mark the maximum boundary of the foreground wedge, the foreground power that reaches this threshold could be much lower than the thermal noise level.
\paragraph{Impact of $u\varv$ coverage on source lines:}The source delay line equations have been derived assuming circular $u\varv$ tracks. However, this assumption breaks down for phase centers far away from the NCP or the SCP, particularly for latitudes close to the equator where the $u\varv$ tracks can deviate considerably from being circular. We have seen in the simulations that the source line equations still describe these situations in general. However, in specific situations where the $u\varv$ tracks deviate significantly from being circular, the signature of the source might not be well described by the derived equations. This is investigated in Appendix~\ref{sec:source_line_specific} for a specific situation where the $u\varv$ coverage has radial structures in the direction of the source. The maximum and minimum of the source line equations shown in Figs.~\ref{fig:source} and \ref{fig:source_appx} do not necessarily indicate the region in which the power will be confined in the parameter space. This is because, in the derivation of the equation, we assume that the maximum power in the $u\varv$ plane occurs exactly along a line in the direction of the source. In reality, this is not always the case even for circular $u\varv$ tracks. This can result in power spilling slightly beyond the minimum and maximum source lines for the observation duration, particularly for phase centers close to the physical horizon. This is discussed in more detail at the end of Sect.~\ref{sec:sim_source}.
\paragraph{Non-ideal basis:}The horizon and source line equations derived in this analysis describe the impact of foregrounds on the cylindrical power spectrum constructed from gridded data in the $u\varv\eta$ domain. However, this Cartesian Fourier domain is not the ideal basis for wide-field instruments. Instead, a spherical harmonic power spectrum, with a spherical spatial Fourier basis, is more appropriate here \citep[e.g.,][]{liu2016spherical, ghosh2018deconvolving}. A cylindrical power spectrum is usually employed by 21-cm cosmology experiments because the $u\varv\nu$ domain is a natural basis for interferometers. It should be noted that even for instruments with a narrow primary beam FoV, where the cylindrical power spectrum is appropriate for describing the 21-cm signal, foreground sources far from the phase center might still be visible in the primary beam sidelobes. As a result, curved sky effects still need to be taken into account, either by using the equations derived in this paper or by switching to a spherical spatial basis.

\subsection{Features and practical applications}
The equations derived in this analysis have several important features that give us new insights into the nature of foreground contamination in the power spectrum. Additionally, there are several situations where the equations derived in this analysis will significantly aid our objective of detecting the 21-cm signal.
\paragraph{Dependence of the foreground wedge on observation parameters:} The conventional horizon and source line equations suggest that the foreground wedge and the signature of strong sources in the cylindrical power spectrum are independent of time, latitude, and the location of the phase center. We show that this is not the case if we consider foregrounds across the full sky, and the signature of the foregrounds depends on several parameters. The horizon line can even be vertical in the extreme example of the phase center crossing the physical horizon (Sect. \ref{sec:ps_full_synthesis}), and in such cases, the foreground wedge covers the entire ($k_{\perp},k_{\parallel}$) space. For the source line, the dependence on time results in a wedge-like signature in the cylindrical power spectrum for a synthesis observation. The derived equations (Eqs. (\ref{eq:horizon_line_lst}), (\ref{eq:horizon_line_full}) and (\ref{eq:source_line})) are presented as analytical functions of the observation parameters like the RA, Dec of the phase center, LST, and the latitude of the interferometer, making them directly applicable to actual observations.
\paragraph{Generalized version of conventional equations:} The conventional equations for both the horizon line (Eq.~(\ref{eq:wedge_flat})) and the source line (Eq.~(\ref{eq:source_line_flat})) are special cases of the equations derived in this paper. Both the LST-dependent horizon line equation and the source line equation reduce to these conventional equations for zenith phasing (snapshot drift scan observations) where it does not matter whether we impose a flat-sky approximation. Therefore, the equations derived in this analysis serve as a more general version of the conventional equations.
\paragraph{Impact on foreground avoidance approach:} Most radio interferometers aiming for a statistical detection of the 21-cm power spectrum are phased arrays. These are often also wide-field instruments that are sensitive to sources close to the physical horizon. Thus, the cylindrical power spectra estimated from their data will have foreground power well beyond the conventional horizon line extent. Foreground avoidance analyses involve the selection of an ``EoR window'' that is assumed to be free of foregrounds. The updated horizon line equations derived in this paper describe the actual maximum extent of foregrounds and will allow a more prudent choice of the ``EoR window".
\paragraph{Impact on foreground subtraction approach:} In foreground subtraction approaches with Gaussian process regression \citep{mertens2018statistical,mertens2024retrieving,acharya202421}, it is necessary to provide priors on the expected frequency coherence scale of foregrounds, to separate them from the 21-cm signal. These new horizon line equations will allow us to provide a more accurate prior on the frequency coherence scale as a function of $k_{\perp}$.
\paragraph{Planning observations to minimize the wedge:} The horizon line equations suggest that the extent of the foreground wedge is closely related to the location of the phase center. This understanding will help in planning observations so that the extent of the horizon wedge is minimized. Planning observations such that the phase center is always above a certain altitude will decrease the extent of the foreground wedge at the cost of lower integration speeds due to limited time spent observing the field. The ideal trade-off will depend on the foreground mitigation strategy used, the modes of interest, and the thermal noise sensitivity of the instrument compared to expected 21-cm signals.
\paragraph{Distinguish between foregrounds and systematics:} The source line equations will allow us to more accurately diagnose the impact of bright sources on the cylindrical power spectrum and correctly distinguish between features in the power spectrum caused by instrumental systematics and those caused by bright sources. Features that lie beyond the conventional horizon lines and were previously, maybe erroneously, assumed to be caused by instrumental or data processing effects could instead be caused by strong foreground sources far from the phase center.
\paragraph{Impact of bright sources on the $u\varv$ plane:} The understanding of the origin of the signature of sources in the gridded $u\varv$ plane can help pinpoint the range of modes most strongly affected by a bright source. This can help in systematically flagging these sources in gridded data accurately. It could also be utilized to subtract the power due to foregrounds utilizing the intrinsic frequency smoothness of foregrounds. Additionally, it can serve as a diagnostic to distinguish between power caused by foregrounds and that caused by instrumental effects or radio frequency interference.
\paragraph{Planning observations to minimize the impact of strong sources:} We have seen that in certain situations the source lines in the cylindrical power spectrum can be extremely steep. The maximum steepness is caused by sources lying on the physical horizon, in the direction opposite to the phase center. This steepness poses considerable challenges since it restricts access to the smallest $k$ modes or largest scales where we have the highest sensitivity for foreground avoidance approaches. For foreground subtraction approaches, steep source lines make blind separation of foregrounds from the 21-cm signal difficult since the foregrounds now exhibit strong spectral fluctuations and are more difficult to separate from the 21-cm signal. The source line equations derived in this analysis can help plan observation phase centers or LST ranges in a way that the bright sources do not result in steep source lines which can extend to high $k_{\parallel}$ values.

\section{Summary}\label{sec:summary}
In this paper, we derive and generalize the equations that describe the full-sky foreground wedge in the cylindrical power spectrum for 21-cm cosmology analyses. These equations are derived without taking a flat-sky approximation and accurately describe the signature of foregrounds in the cylindrical power spectrum for an arbitrary phase center. We derive two kinds of equations, corresponding to the horizon and source lines. Horizon lines define the maximum extent of foregrounds in the cylindrical power spectrum, whereas source lines describe the signature of a specific source. Both sets of equations are tested against full simulations for a fiducial interferometric array placed at the locations of NenuFAR, LOFAR, and SKA, across a range of observational parameters. We find that the foreground wedge can extend well beyond what the conventional horizon line equations suggest, and the equations derived in this paper perfectly describe this extended horizon limit. The updated source line equations significantly improve the characterization of the impact of a specific strong source on the cylindrical power spectrum. These new equations are crucial for correctly describing the signature of foregrounds in 21-cm cosmology analyses employing either foreground avoidance or foreground subtraction strategies. Additionally, the equations will allow a more informed choice of the phase center and LST ranges of observations, particularly in the context of wide-field instruments such as the NenuFAR, LOFAR, MWA, and the future SKA, that employ the reconstructed power spectrum approach.

We have made a Python library \texttt{pslines}\footnote{\url{https://github.com/satyapan/pslines}} openly available. This generic library can generate and plot the horizon and source lines using the new equations presented in this paper, for any tracking interferometer.

\begin{acknowledgements}
SM, LVEK, SAB, JKC, BKG, SG, and CH acknowledge the financial support from the European Research Council (ERC) under the European Union’s Horizon 2020 research and innovation programme (Grant agreement No. 884760, "CoDEX”). FGM acknowledges the support of the PSL Fellowship. EC would like to acknowledge support from the Centre for Data Science and Systems Complexity (DSSC), Faculty of Science and Engineering at the University of Groningen.
\end{acknowledgements}

\bibliographystyle{aa}
\bibliography{aa}

\begin{appendix}
\section{Maximum delays for NCP phasing}\label{sec:ncp_phasing}
The NCP is the main target field for the LOFAR and NenuFAR 21-cm projects. Here we analytically derive the expression for the maximum delay for NCP phasing, by maximizing the expression for delay (Eq.~(\ref{eq:delay_arbitrary})) subject to the horizon condition (Eq.~(\ref{eq:horizon_mask_arbitrary})). We work in the XYZ coordinate system described in Sect.~\ref{sec:analytical} and set $\delta_0=90^{\circ}$ and $\phi \geq 0$ for NCP phasing. To make the problem analytically tractable, we reduce it to the meridian plane. It should be noted that this can only be done without loss of generality when the phase center is at the NCP or SCP since these are the only two points in the sky that are fixed relative to the Earth. The baseline vector and source responsible for the maximum delay lie on the meridian plane. We choose a baseline pointing toward the north and a source toward the south to get the extreme case of maximum absolute delay. This amounts to setting $H=0^{\circ}$ and $\theta=90^{\circ}$ in Eqs. (\ref{eq:delay_arbitrary}) and (\ref{eq:horizon_mask_arbitrary}). Thus, we need to maximize the equation:
\begin{equation}\label{eq:delay_ncp}
\eta_0(\phi, \delta) = \dfrac{|\vec{b}|}{c}\, \Bigl(-\sin\phi \cos\delta + \cos\phi \sin\delta - \cos\phi\Bigr),
\end{equation}
subject to the horizon condition:
\begin{equation}\label{eq:horizon_mask_ncp_2d}
\delta \geq \phi-\frac{\pi}{2}.
\end{equation}
The maximum absolute delay is now analytically obtained by setting $\frac{\partial \eta_0}{\partial \delta} = 0$. This gives us:
\begin{equation}\label{eq:horizon_ncp}
\delta\Bigr|_{\mathrm{max}} = \phi - \dfrac{\pi}{2} \text{ ; } |\eta_0|^{\mathrm{max}} = \dfrac{|\vec{b}|}{c}\Bigl(1+\cos\phi\Bigr).
\end{equation}
This extremum also satisfies the horizon condition of Eq.~(\ref{eq:horizon_mask_ncp_2d}) and the source lies exactly on the physical horizon, as was seen for the case of an arbitrary phase center (Sect.~\ref{sec:max_delays}). To ensure that this reduction of the problem to the meridian plane can indeed be made without loss of generality, Eq.~(\ref{eq:delay_arbitrary}) was numerically maximized subject to Eq.~(\ref{eq:horizon_mask_arbitrary}) over $H$, $\delta$ and $\theta$ for $\delta_0=90^{\circ}$, $|\vec{b}| = 100$ and $\phi = 47.38^{\circ}$. The maximum occurs at $(H, \delta, \theta) \approx (0, -42.62^{\circ}, 270^{\circ})$ with a value of $5.59\times 10^{-7}\,$s while the minimum occurs at $(H, \delta, \theta) \approx (0, -42.62^{\circ}, 90^{\circ})$ with a value of $-5.59\times 10^{-7}\,$s. Both situations occur at a declination of $\phi-\pi/2$ and have a delay value of $(|\vec{b}|/c)(1+\cos\phi)$. The maximum occurs when the baseline points along the south and the minimum occurs when the baseline points along the north, as is expected for a source located toward the south. These results match with the predictions of Eq.~(\ref{eq:horizon_ncp}) (irrespective of whether Eq.~(\ref{eq:horizon_mask_ncp_2d}) is imposed). This shows that for the specific case of NCP phasing, the problem can be reduced to the meridian plane without loss of generality.

For drift scan observations, the horizon delay is given by $|\vec{b}|/c$ which occurs due to sources $90^\circ$ away from the zenith. However, once the data is phased to the NCP (or any other location), we do not need to go  $90^{\circ}$ away from the phase center to get a delay equal to $|\vec{b}|/c$. Setting $|\eta_0| = |\vec{b}|/c$ in Eq.~(\ref{eq:delay_ncp}), we get $\delta = \sin^{-1}(\cos\phi-1)+\phi=28.54^{\circ}$ for $\phi=47.38^{\circ}$ (NenuFAR). So a source located more than $61.46^{\circ}$ away from the phase center already will have a delay greater than $|\vec{b}|/c$. For the horizon line in the power spectrum, we need to set $|\eta_0|/\sin\phi = |\vec{b}|/c$ (following Sect.~\ref{sec:horizon_ps}) where $\sin\phi$ is the projection of $\vec{b}$ in the plane perpendicular to $\hat{\vec{p}}$. The angle $\zeta$ away from the phase center beyond which the horizon line is steeper than the conventional horizon line is then given by:
\begin{equation}\label{eq:gamma_phi}
\zeta = \pi/2-\delta = \pi/2-\sin^{-1}(\cos\phi-\sin\phi)-\phi.
\end{equation}
Inserting $\phi=47.38^{\circ}$ for NenuFAR, we get $\zeta=45.99^{\circ}$. Therefore, the primary beam must be able to attenuate all power beyond an angle less than $45.99^{\circ}$, otherwise the foreground wedge will be steeper than that predicted by the conventional horizon line equation (Eq.~(\ref{eq:wedge_flat})). Since Eq.~(\ref{eq:gamma_phi}) is a monotonically increasing function of $\phi$, this effect is more pronounced for lower latitudes, where sources slightly away from the phase center can result in a foreground wedge steeper than the conventional horizon extent.

\section{Sampling and gridding on a circular $u\varv$~track}\label{sec:sampling_gridding_analytical}
In Sect.~\ref{sec:sampling_gridding}, we find that sampling visibilities along $u\varv$ tracks and gridding them using a convolution kernel results in higher power along the source direction, particularly for sources far away from the phase center. Here we investigate this effect analytically, for a circular track in the $u\varv$ plane with a radius $u_{0}$, on which the interferometer samples the visibility. For a unit unpolarized source $\vec{s'} = (l,m)$, under the flat-sky approximation, the visibility function is given by:
\begin{align*}
V(u,\varv) = e^{-2\pi i(\vec{u}\cdot\vec{s'})} = e^{-2\pi i (ul+\varv m)} \text{ where }\vec{u}=(u,\varv).
\end{align*}
The sampled visibility on the circular track is given by:
\begin{align*}
    V_{\mathrm{track}}(u,\varv) &= e^{-2\pi i (ul+\varv m)}  S(u,\varv),\\ \text{where } &S(u,\varv) = \begin{cases} 1 & \mbox{if }\sqrt{u^2+\varv^2} = u_0\\0 & \mbox{otherwise}.\end{cases}
\end{align*}
To reduce the problem to one dimension, we can write down the expression of $V_{\mathrm{track}}$ in terms of $\beta$ (the angle between $\vec{u}$ and $\vec{s'}$).
\begin{align}\label{eq:v_track}
    V_{\mathrm{track}}(\beta) &= e^{-2\pi i |\vec{u}||\vec{s'}|\cos\beta} = e^{-2\pi i u_0 \sin\psi \cos\beta}.
\end{align}
These sampled visibilities are finally distributed on the $u\varv$ plane using a convolution kernel. We assume, for simplicity, that a two-dimensional Gaussian is used as the convolution kernel. If $f$ is the FoV or the width of the Gaussian in the image space, the corresponding gridding kernel in the $u\varv$ space is given by:
\begin{align*}
    G(u,\varv) = e^{-2\pi f^2 (u^2+\varv^2)},
\end{align*}
and the convolved visibility is:
\begin{align*}
    V_{\mathrm{conv}}(u,\varv) &= \int_{u'\varv'}V_{\mathrm{track}}(u',\varv')G(u-u',\varv-\varv')du'd\varv'\\
    &= \int_{u'\varv'}V_{\mathrm{track}}(u',\varv')e^{-2\pi f^2|\vec{u}-\vec{u'}|^2}  du'd\varv'.
\end{align*}
Now $V_{\mathrm{track}}(u',\varv')$ is zero everywhere except where $\sqrt{u^2+\varv^2}=u_0$. On such a track, $\vec{u}$ and $\vec{u'}$ form two sides of an isosceles triangle with the length of the equal sides being $u_0$ and the angle between the sides being $\beta-\beta'$. Thus, the length of the third side ($|\vec{u}-\vec{u'}|$) is $u_0\sqrt{2-2\cos(\beta-\beta')}$. Now since the convolution does not get any contribution from visibilities outside the circular track, the expression for the convolved visibility along the track becomes:
\begin{align*}
    V_{\mathrm{conv}}(\beta) &= \int_{\beta'}V_{\mathrm{track}}(\beta')e^{-4\pi f^2 u_0^2 (1-\cos(\beta-\beta'))}u_0 d\beta'\\
    &= V_{\mathrm{track}}(\beta)\otimes G(\beta),\text{ where } G(\beta) = u_0 e^{-4\pi f^2 u_0^2 (1-\cos\beta)}.
\end{align*}
Here $u_0 d\beta'$ is an infinitesimal element along the track and `$\otimes$' denotes a convolution over $\beta$. Finally, inserting the expression of $V_{\mathrm{track}}(\beta)$ from Eq.~(\ref{eq:v_track}), we get:
\begin{align}\label{eq:v_conv}
    V_{\mathrm{conv}}(\beta) &= \cos(2\pi u_0 \sin\psi \cos\beta)\otimes G(\beta) \nonumber\\&+ i\sin(2\pi u_0 \sin\psi \cos\beta)\otimes G(\beta).
\end{align}
\begin{figure}
    \includegraphics[width=\hsize]{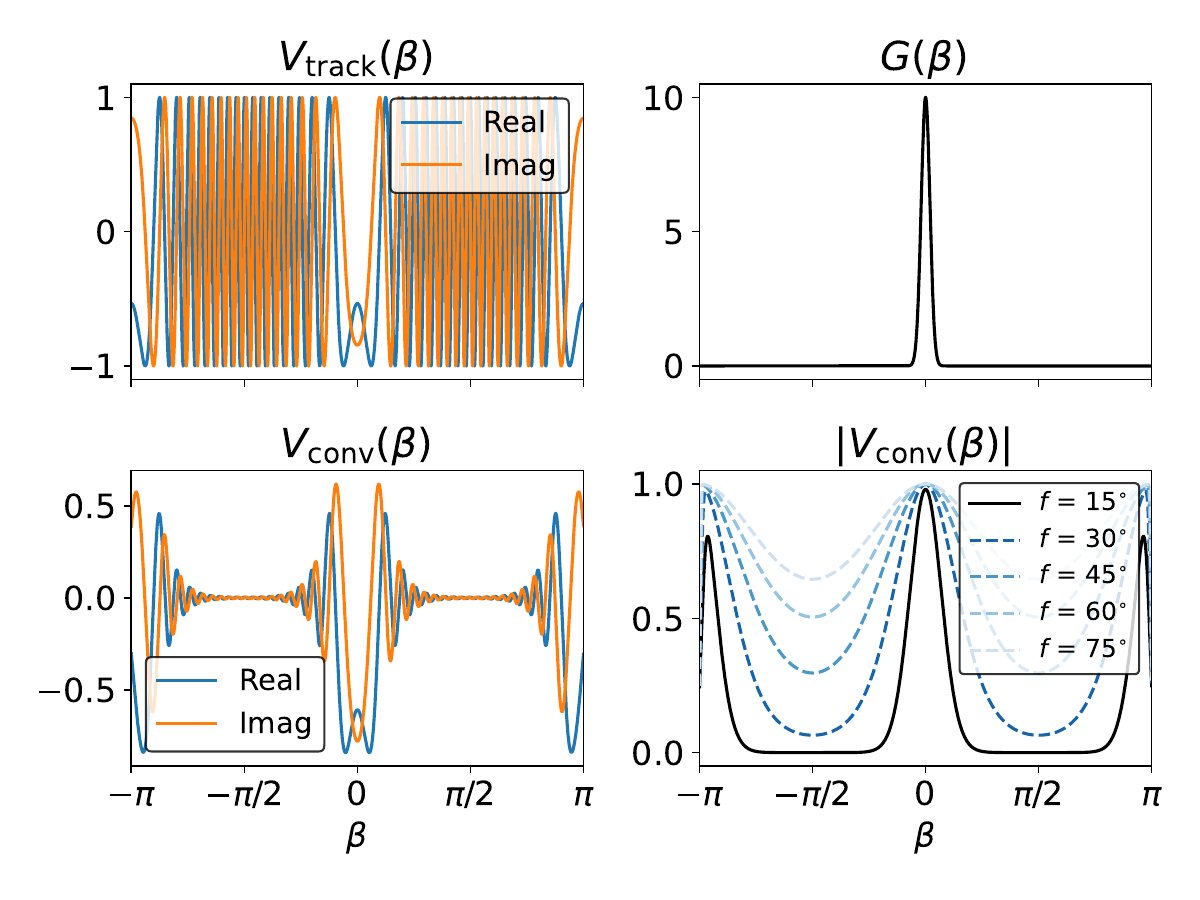}
    \caption{The impact of sampling visibilities due to a source $60^{\circ}$ away from the phase center ($\psi = 60^{\circ}$) on a circular $u\varv$ track of radius 10 ($u_0=10$) followed by convolution with a gridding kernel corresponding to a FoV of $15^{\circ}$ ($f=15^{\circ}$). The different panels show the visibilities along the $u\varv$ track (top-left, Eq.~(\ref{eq:v_track})), the convolution kernel (top-right), the sampled visibilities convolved with the kernel (bottom-left, Eq.~(\ref{eq:v_conv})) and the absolute value of the convolved visibilities (bottom-right). The amplitudes of the convolved visibilities for a few other FoVs are overplotted on the bottom-right panel in blue dashed lines.}
    \label{fig:source_signature_1d}
\end{figure}

Figure~\ref{fig:source_signature_1d} illustrates the effect of convolving $V_{\mathrm{track}}(\beta)$ with $G(\beta)$ for $u_0=10$, $\psi = 60^{\circ}$ and $f=15^{\circ}$. Both the real and imaginary parts of $V_{\mathrm{track}}(\beta)$ are quasi-sinusoids that have fast oscillations near $\beta=-\pi/2$ and $\beta=\pi/2$ (top-left panel). These fast oscillations are suppressed by convolution with the gridding kernel (top-right panel) in both real and imaginary parts of $V_{\mathrm{conv}}(\beta)$ (bottom-left panel). As a result, the amplitude of $V_{\mathrm{conv}}(\beta)$ is lowered near $\beta=-\pi/2$ and $\beta=\pi/2$ (bottom-right panel). The bottom-right panel also shows the effect of the FoV on the amplitude of the convolved visibilities. Larger FoVs (narrower gridding kernels) filter out less power and result in a wider signature. This is also seen in the middle and right panels of Fig.~\ref{fig:source_signature_2d}. Changing the radius of the $u\varv$ track ($u_0$) does not change the width of the peak of $|V_{\mathrm{conv}}(\beta)|$ (not shown in the figure). This is because, even though the fluctuations of $|V_{\mathrm{track}}(\beta)|$ are faster in a longer $u\varv$ track, $G(\beta)$ is also narrower (since the same $G(u,\varv)$ is now evaluated on a longer $u\varv$ track), thus exactly compensating for the effect. This means that the signature of a source in the $u\varv$ plane is a sector with a higher power, and the FoV decides its opening angle $\Delta\beta$. This is seen in the middle and right panels of Fig.~\ref{fig:source_signature_2d}. Also, the nature of the convolution kernel affects the exact shape of $|V_{\mathrm{conv}}(\beta)|$. Thus, for non-Gaussian convolution kernels, the power need not be the highest at exactly $\beta=0$. It should be noted that here we have assumed a single $u\varv$ track. For multiple $u\varv$ tracks, Eq.~(\ref{eq:v_conv}) is not exactly valid since the convolved visibility at a point on a given $u\varv$ track will have a contribution from other $u\varv$ tracks. Also, we have assumed the $u\varv$ track is circular. Thus, the track runs perpendicular to the oscillations exactly at $\beta=n\pi$ (for integer n). At these points, both the real and imaginary parts of the visibility have the slowest fluctuations along the track. As a result, the convolution does not decrease the amplitude of the visibilities at $\beta=n\pi$, leading to higher power in the source direction. For a non-circular track, the slowest oscillations along the track do not necessarily occur exactly at $\beta=n\pi$. Thus, the signature of the source on the $u\varv$ plane will be shifted accordingly, and cannot be predicted without knowing the exact $u\varv$ coverage of the observation.

\section{Source delay validation}\label{sec:source_delay_validation}
Given a specific source, the delay corresponding to the maximum power in the power spectrum is given by Eq.~(\ref{eq:source_delay}). The baseline for which we get this delay lies along the projection of $\hat{\vec{s}}-\hat{\vec{p}}$ on the Earth's tangent plane. Incidentally, this is also the baseline orientation which results in the maximum delay, since it makes the smallest angle with the $\hat{\vec{s}}-\hat{\vec{p}}$ vector. As a result, this expression of delay not only corresponds to the situation of maximum power but also the maximum delay. We verify this by maximizing Eq.~(\ref{eq:delay_arbitrary}) over $\theta$ for Cyg A and Vir A, corresponding to $t=18\,$h.
\begin{figure}
    \includegraphics[width=0.48\hsize]{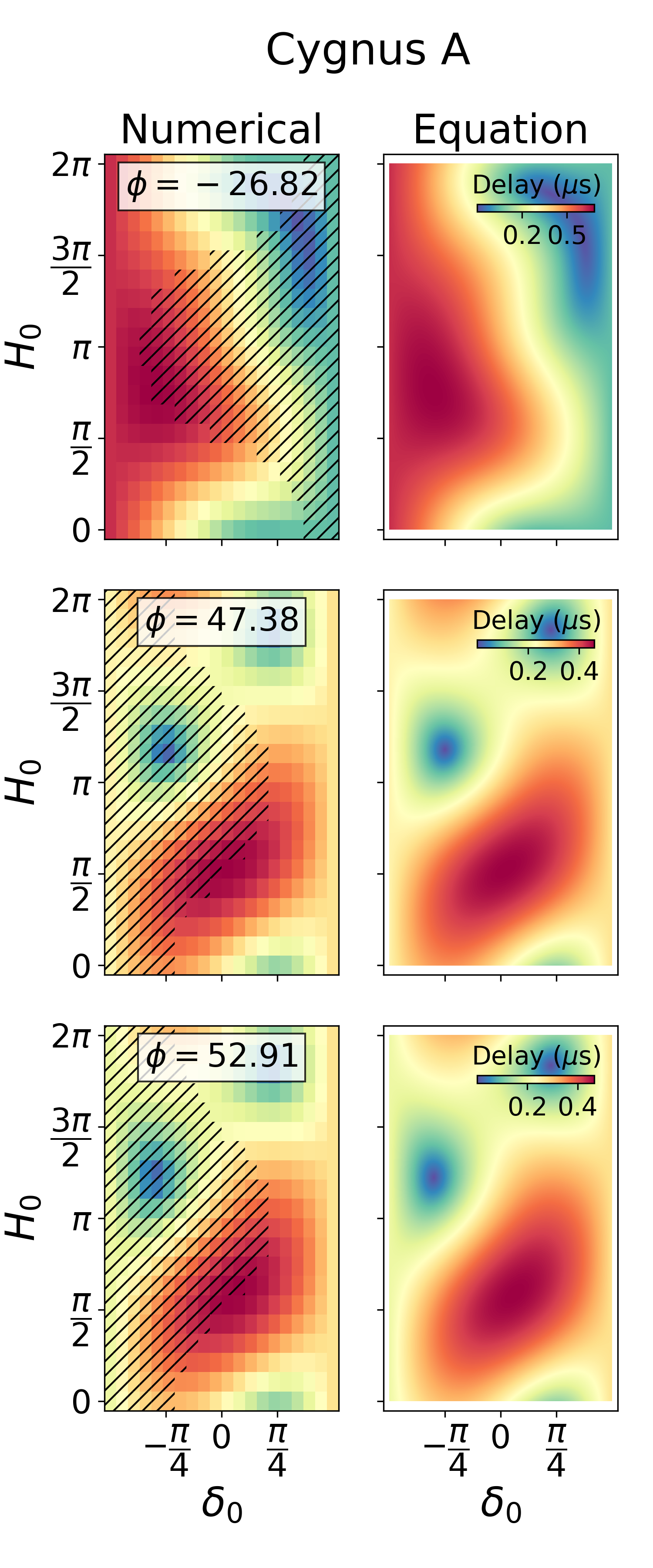}
    \includegraphics[width=0.48\hsize]{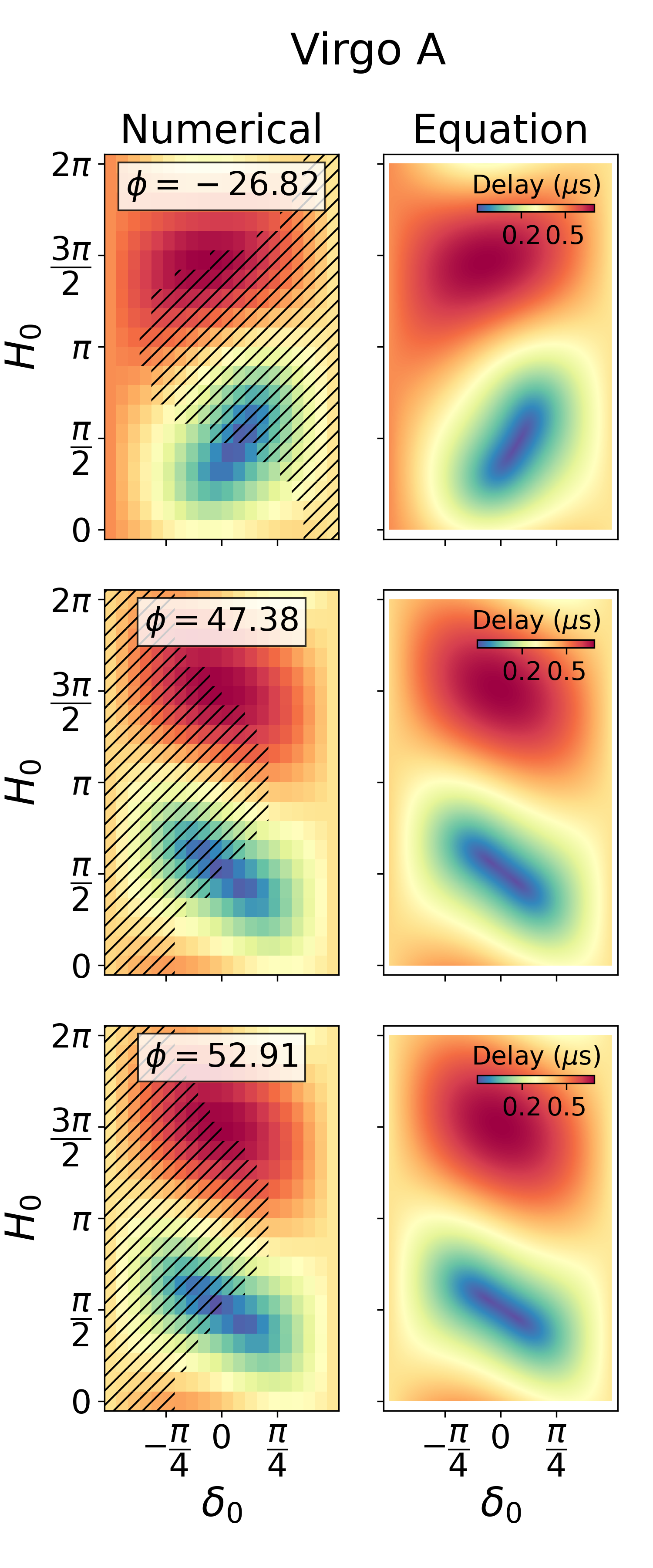}
    \caption{Comparison between the delay due to a source obtained by numerical maximization of Eq.~(\ref{eq:delay_arbitrary}), against that expected from Eq.~(\ref{eq:source_delay}). The delays due to Cyg A (left two columns) and Vir A (right two columns) for three different latitudes (different rows) are shown. The columns titled ``Numerical'' correspond to delay values obtained by numerically maximizing Eq.~(\ref{eq:delay_arbitrary}) over $\theta$. The columns titled ``Equation'' correspond to the delay values for the baseline responsible for maximum power as predicted by Eq.~(\ref{eq:source_delay}). The slanted stripes indicate regions where the phase center is below the physical horizon.}
    \label{fig:source_delay_cyga_vira}
\end{figure}
Fig.~\ref{fig:source_delay_cyga_vira} shows how the numerically maximized delays due to Cyg A and Vir A compare against those predicted by Eq.~(\ref{eq:source_delay}), for all possible phase center positions. We see that the delays match for both sources and for all three latitudes.

\section{Source lines for specific cases}\label{sec:source_line_specific}
\begin{figure}
    \includegraphics[width=\hsize]{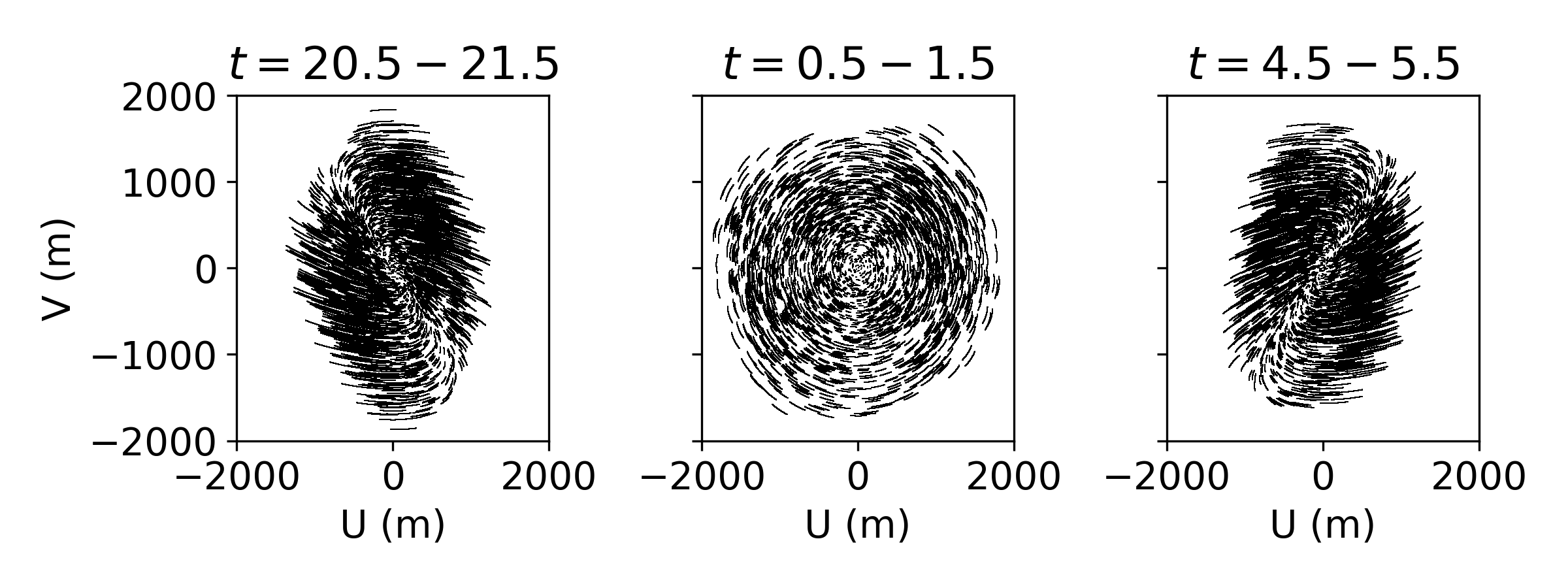}
    \includegraphics[width=\hsize]{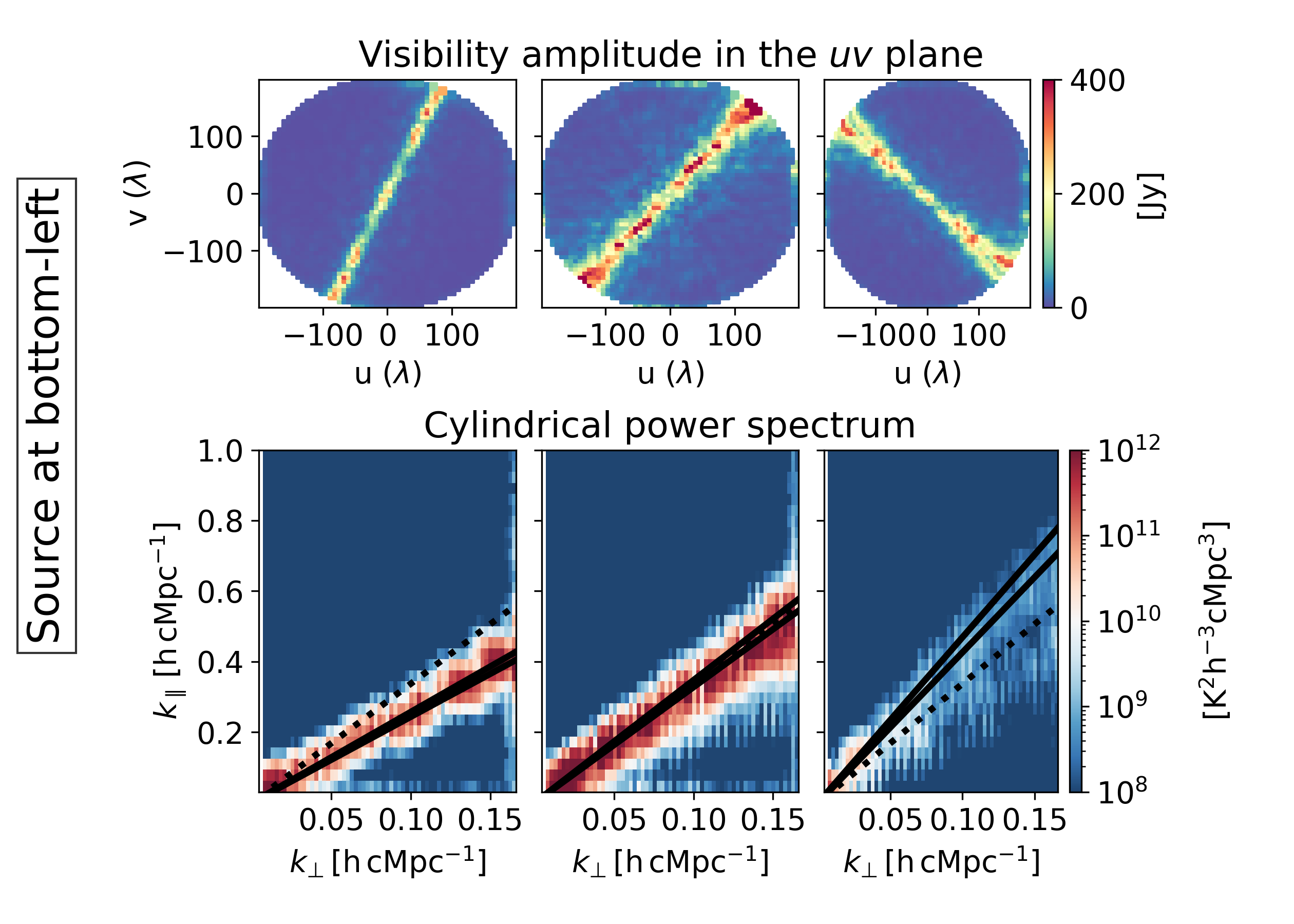}
    \includegraphics[width=\hsize]{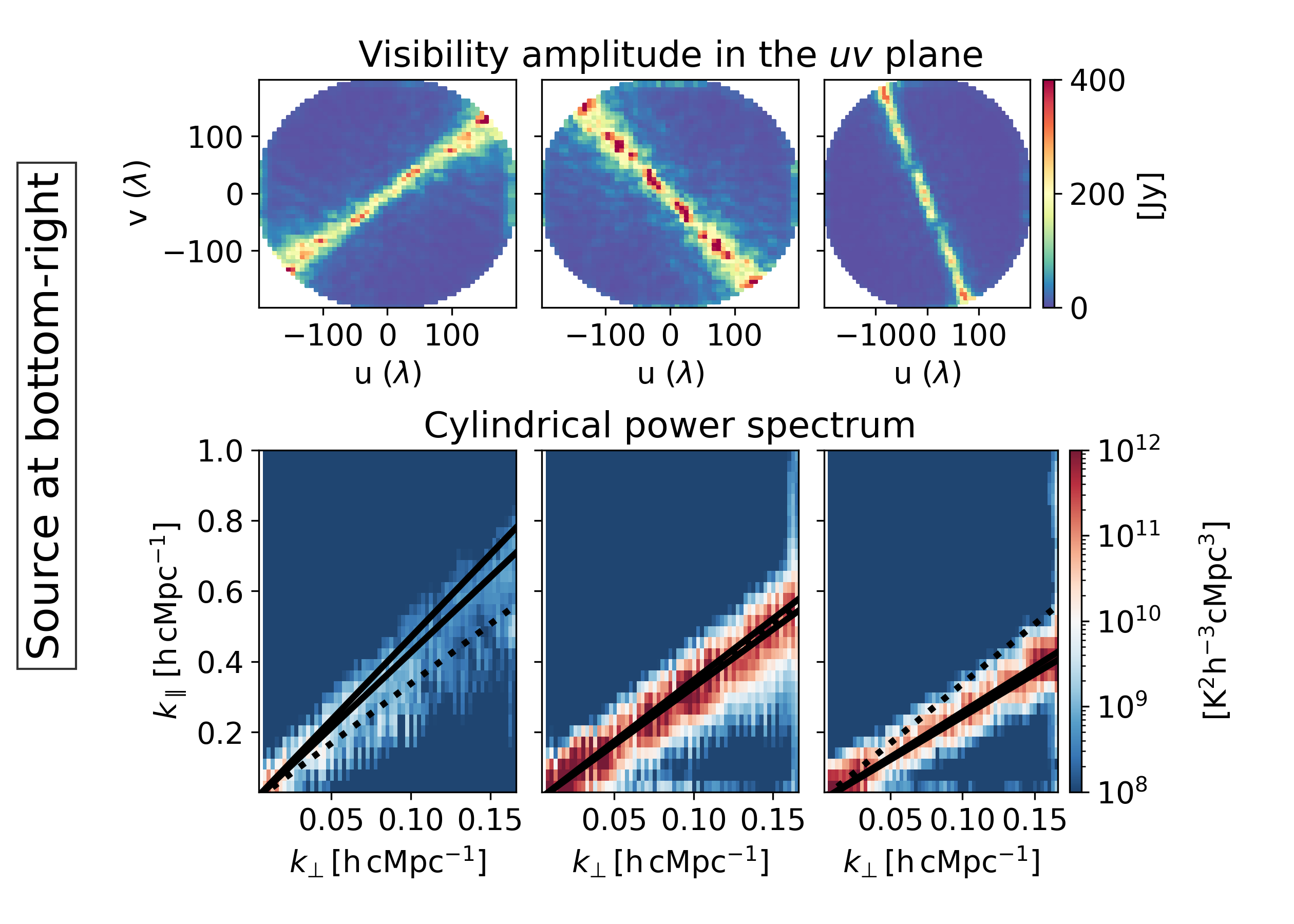}
    \includegraphics[width=\hsize]{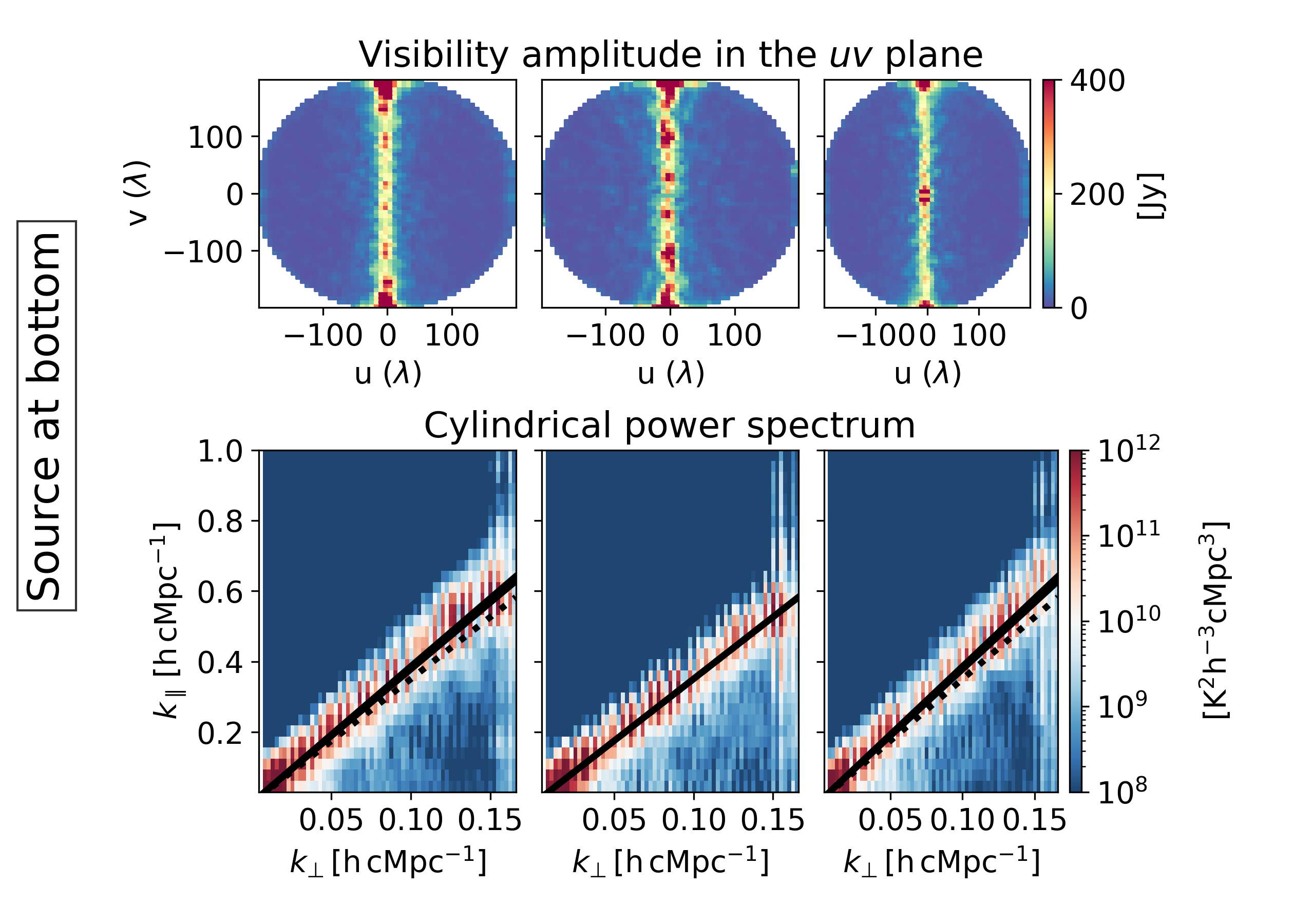}
    \caption{The impact of $u\varv$ coverage on source lines. The top row shows the baseline coverages for the three LST ranges for which the simulations are repeated. The other rows show the visibility amplitude in the $u\varv$ plane and the cylindrical power spectra, for a source located at the bottom-left, bottom-right, and bottom of the image.}
    \label{fig:radial_features}
\end{figure}

In Sect.~\ref{sec:sampling_gridding} we find that sampling on $u\varv$ tracks and gridding with a convolution kernel results in the visibilities due to a source having a specific signature in the $u\varv$ plane. In the derivation of the source line equation (Eq.~(\ref{eq:source_line})), we assumed that the maximum power in the $u\varv$ plane occurs along the direction of the source, which is exactly correct only for circular $u\varv$ tracks. For telescopes located at latitudes close to the equator, the $u\varv$ tracks can look very different, particularly when the phase center moves far away from the zenith. Here, we investigate the impact of non-circular $u\varv$ tracks on the source lines, in the $u\varv$ plane and the cylindrical power spectrum. We performed simulations with the fiducial array described in Sect.~\ref{sec:simulations} placed at the location of SKA, with a zenith transit phase center (RA = $1\,$h, Dec = $-26.82^{\circ}$). Three $1\,$h simulations centered at LSTs of $21\,$h, $1\,$h, and $5\,$h were performed, so that the phase center crosses the zenith at the midpoint of the simulation centered at LST = $1\,$h. The $u\varv$ coverages for the three simulations are shown in the top row of Fig.~\ref{fig:radial_features}. We can see that the first and third simulations result in radially outward $u\varv$ tracks along the bottom-right and bottom-left respectively. The simulations were repeated for three different sources. The first source chosen is Fornax A ($\alpha=3.4\,$h, $\delta=-37.21^{\circ}$), which lies in the bottom-left corner of the image plane. We then repeated the simulations for a source placed at the same declination, but at $\alpha=22.6\,$h, so that it lies on the bottom-right corner of the image plane. Finally, we repeated the simulations for a source located toward the bottom of the image plane ($\alpha=1\,$h, $\delta=-60^{\circ}$). The visibility amplitudes in the $u\varv$ plane for all simulations are shown in Fig.~\ref{fig:radial_features}. We see that the source located at the bottom-left corner results in higher power approximately along this direction for the first two LSTs, but for the third LST, the line of higher power is in a completely different direction. This happens because for a source on the bottom-left corner, for the third LST, most of the $u\varv$ tracks run along the direction of the source. As a result, there are fast fluctuations of the visibility due to the source along these $u\varv$ tracks, which are suppressed by the convolution kernel used in gridding (as described in Sect.~\ref{sec:sampling_gridding}). The opposite scenario occurs for the simulations with the source on the bottom-right since now the first LST simulation has $u\varv$ tracks running in the direction of the source. For the simulations with the source at the bottom, the source line in the $u\varv$ plane behaves as expected for all three LSTs, since the $u\varv$ tracks do not run along the source direction for any of the three LSTs. To see the impact of the $u\varv$ tracks running in the direction of the source on the cylindrical power spectrum, we filtered out the $u\varv$ cells which lie more than $40^{\circ}$ away from the source direction and constructed the power spectrum. Fig.~\ref{fig:radial_features} shows that when the source is at the bottom-left or the bottom-right of the image plane, the power due to the source is reduced whenever we have $u\varv$ tracks running along the source direction. When the source is at the bottom, this reduction of power does not occur, as we would expect when the $u\varv$ tracks do not run along the source direction. The source line equation derived in this analysis (Eq.~(\ref{eq:source_line}): black solid line) and the conventional source line equation with a flat-sky approximation (Eq.~(\ref{eq:source_line_flat}): black dotted line) are also shown in the plots. We see that the conventional source line equation can accurately predict the signature of the source only when the phase center is at the zenith (as described in Sect.~\ref{sec:drift_scan}), but the new equation can describe the signature in other situations as well.

\section{Additional figures}
\begin{figure*}
    \includegraphics[width=\hsize]{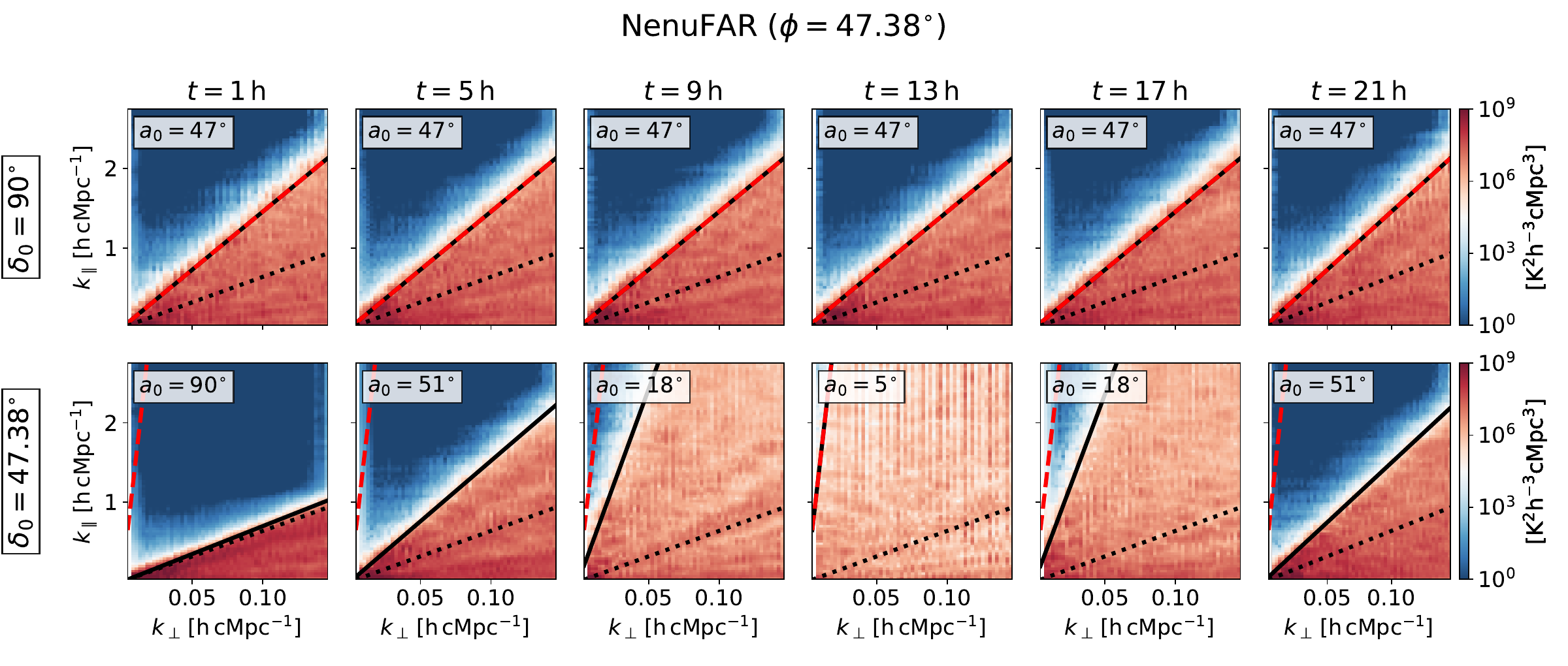}
    \includegraphics[width=\hsize]{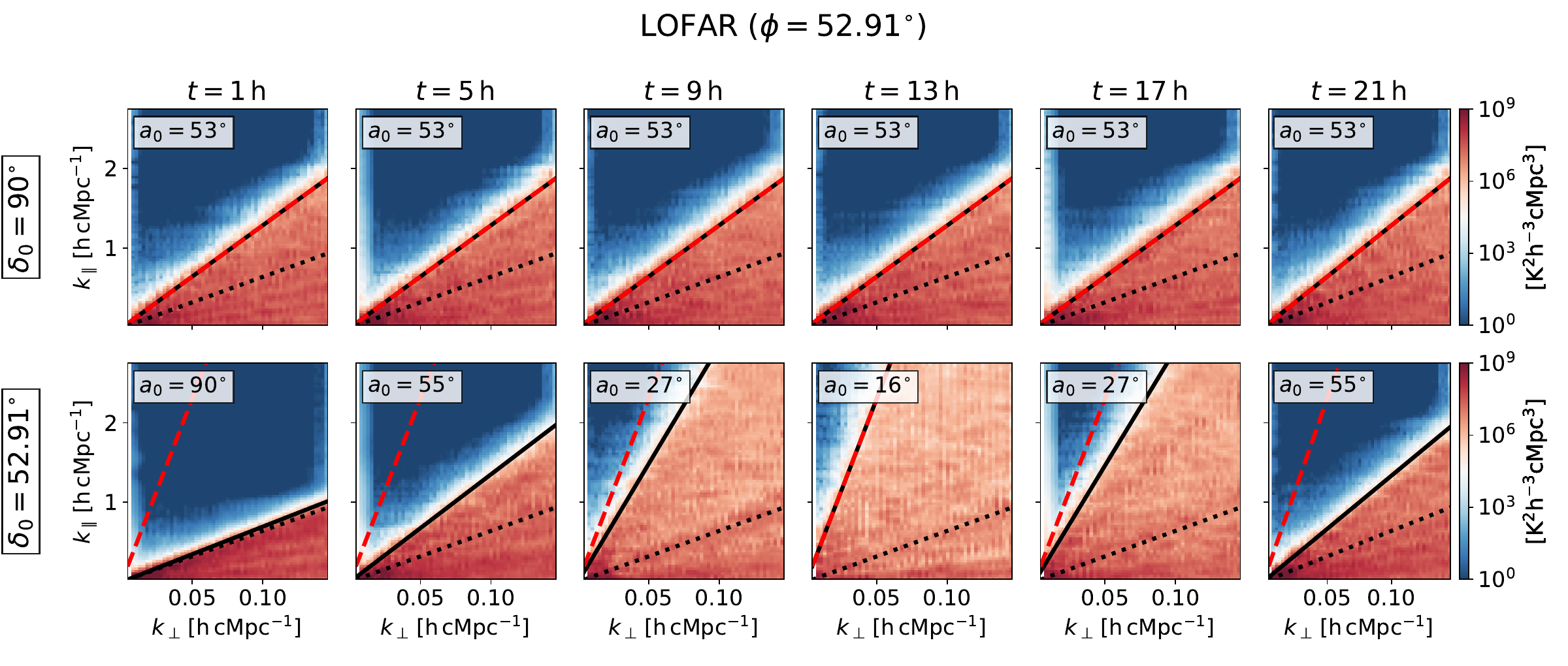}
    \includegraphics[width=\hsize]{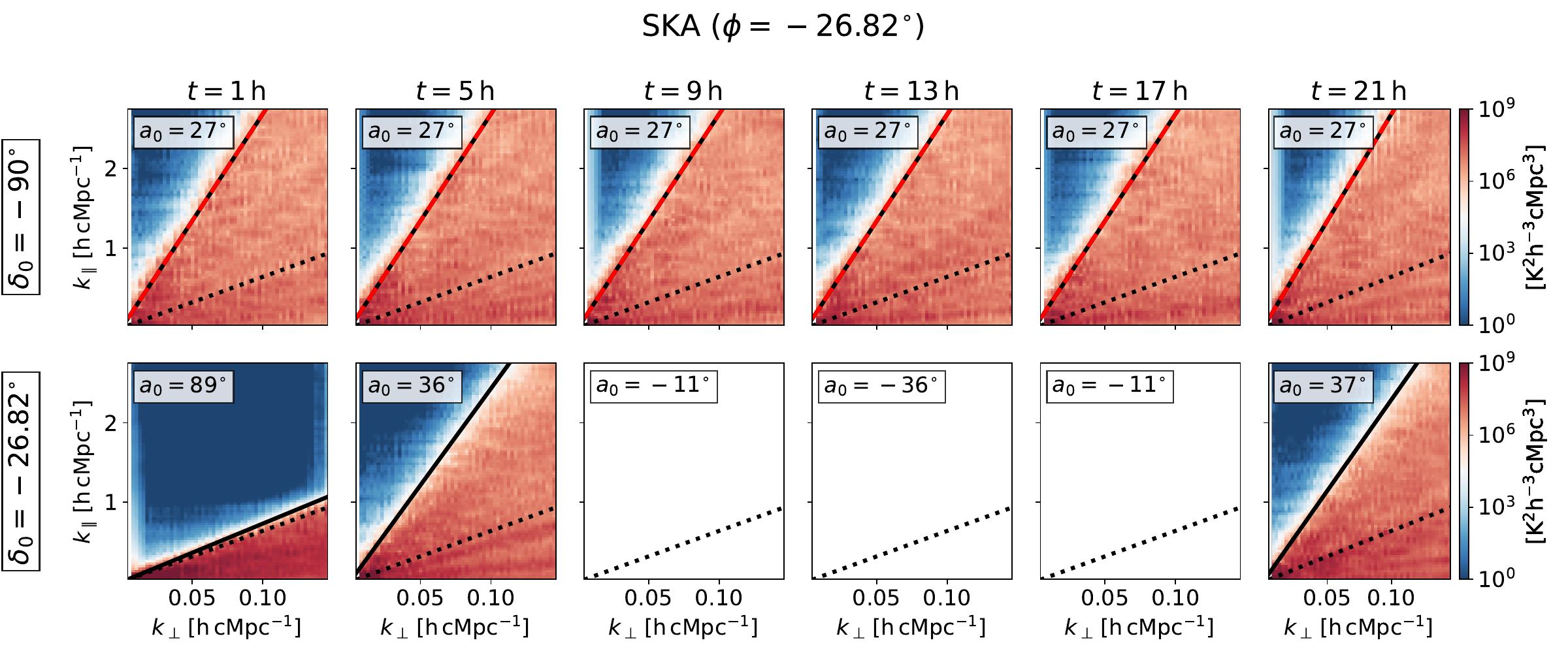}
    \caption{Cylindrical power spectra for the remaining full-sky simulations that were not shown in Fig. \ref{fig:horizon}. For each location, the two rows correspond to the celestial pole (top) and zenith transit (bottom) phase center declinations. In the lowermost row, for the simulations centered at LST = 9h, 13h, and 17h, the phase center is below the horizon for part of the observation and hence the visibilities are not predicted. Also, the full synthesis line is not shown in this case since it is vertical (from Eq.~(\ref{eq:horizon_line_full})).}
    \label{fig:horizon_appx}
\end{figure*}

\begin{figure*}
    \includegraphics[width=\hsize]{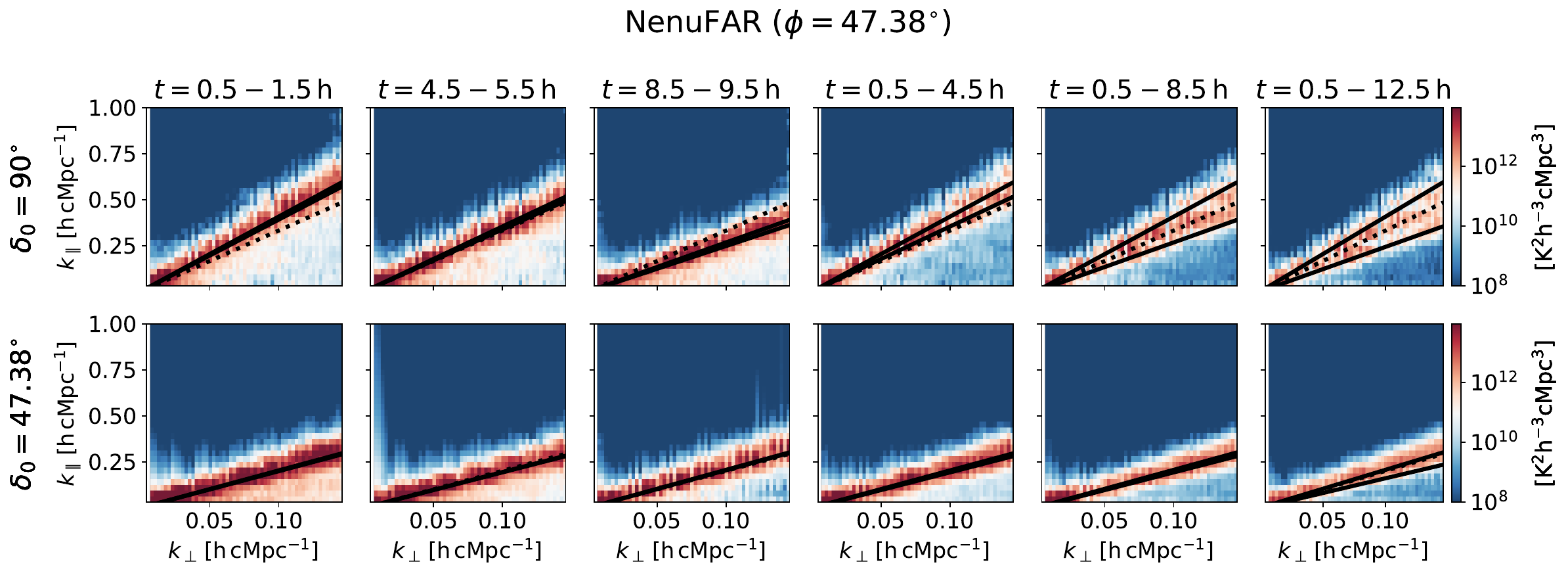}
    \includegraphics[width=\hsize]{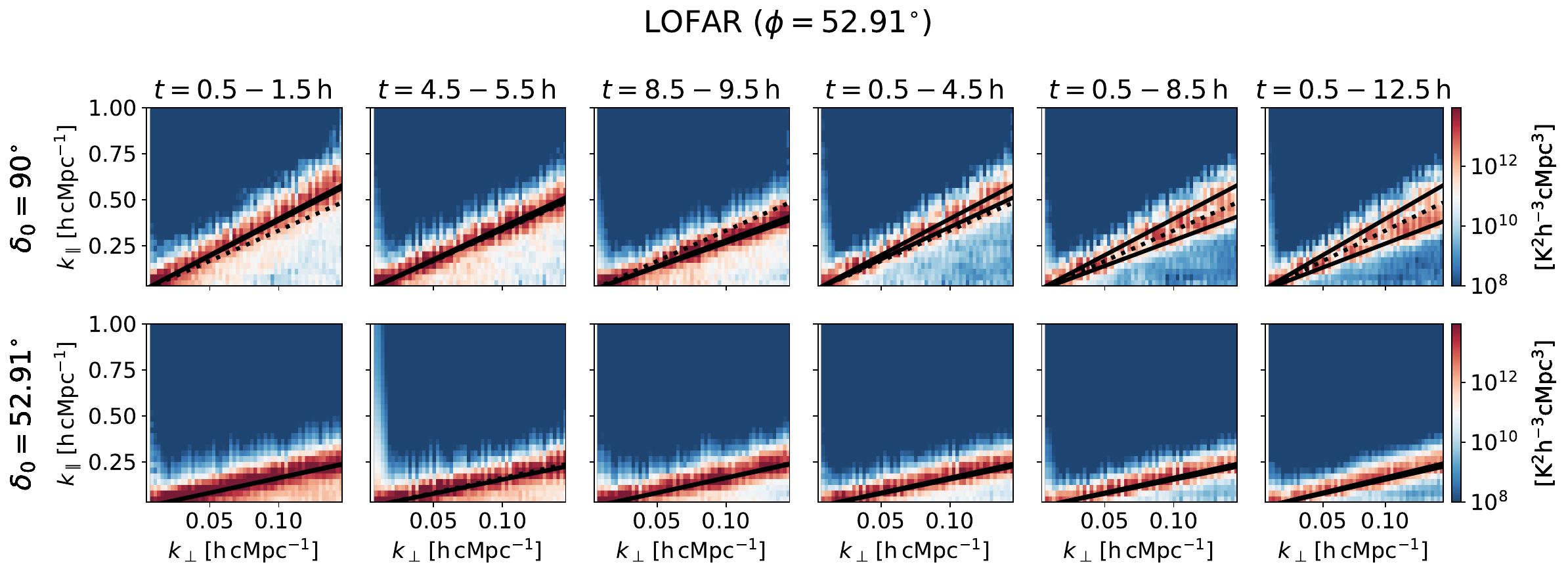}
    \includegraphics[width=\hsize]{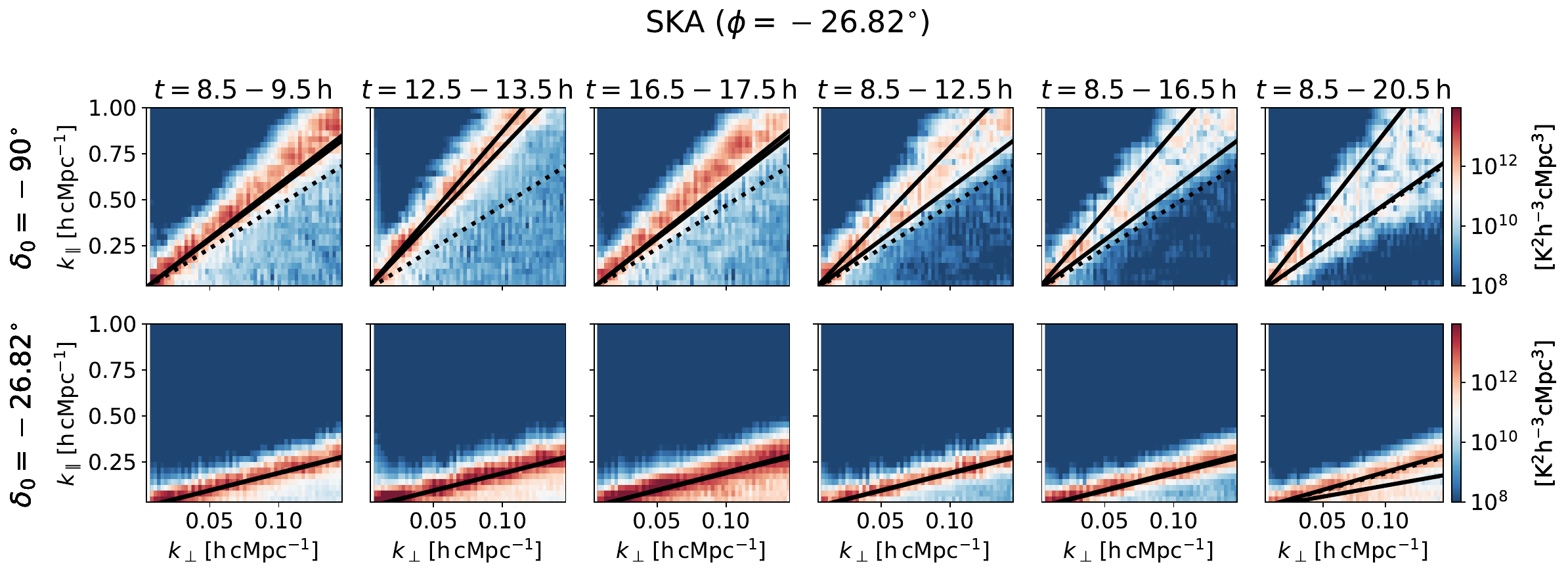}
    \caption{Cylindrical power spectra for the remaining single source simulations that were not shown in Fig. \ref{fig:source}. Here the power spectra correspond to Cas A simulations at the locations of NenuFAR and LOFAR, and Cen A simulations at the location of SKA.}
    \label{fig:source_appx}
\end{figure*}
\end{appendix}

\end{document}